\documentclass[12pt]{article}

\usepackage[utf8]{inputenc}
\usepackage[T1]{fontenc}

\usepackage{amsmath}
\usepackage{amssymb}

\usepackage{graphicx}
\usepackage[dvipsnames]{xcolor}

\usepackage{booktabs}
\usepackage{longtable}

\usepackage{enumitem}

\usepackage[margin=1in]{geometry}

\usepackage[colorlinks=true,linkcolor=blue,citecolor=blue,urlcolor=blue]{hyperref}

\usepackage[round]{natbib}

\usepackage{pdflscape}

\usepackage{setspace}
\usepackage{authblk}

\title{\textbf{Personal Data as a Human Right: A New Social Contract Based on Data Sovereignty, Human Dignity and Data Personalism}}

\author[1]{Alvarez-Pallete, J.M.}
\author[1]{Calder\'{o}n, R.}
\author[1]{Corzo, M.T.}
\author[1,2]{Garrido-Merch\'{a}n, E.C.}
\author[1,2]{L\'{o}pez, G.}
\author[1]{Navarro-Mendiz\'{a}bal, I.}
\author[3]{Padilla, S.}
\author[4]{Pad\'{\i}n, A.}
\author[1]{Redondo, R.}

\affil[1]{Universidad Pontificia Comillas, Madrid, Spain}
\affil[2]{Institute for Research in Technology (IIT), Madrid, Spain}
\affil[3]{Banco de Espa\~{n}a, Madrid, Spain}
\affil[4]{Garrigues, Madrid, Spain}

\date{}

\begin{document}

\maketitle

\begin{abstract}
In an era of ubiquitous data collection, platform dominance, and AI-mediated governance, the implicit social contract of digital life is increasingly shaped by a small number of private actors rather than by democratic deliberation. This position paper advances a dignity-centric Digital Social Contract grounded in data sovereignty, human dignity, and data personalism: the view that personal data are rights-laden emanations of the person and should be protected, in substance, as a human right, not treated as neutral inputs or tradable commodities. Drawing on social contract theory and interdisciplinary scholarship across law, ethics, economics, computer science, sociology, and political philosophy, we diagnose how datafied infrastructures and surveillance-based business models convert everyday traces into profiles, predictions, and consequential decisions at scale, concentrating informational power and weakening meaningful consent, autonomy, and civic trust. Against this backdrop, we contrast DatAIsm (an extractive paradigm that reduces persons to datapoints and optimizes for prediction and control) with HumAIsm, which recenters the human subject and the irreducibility of dignity to mere calculation. We then articulate an operational governance architecture organized around six interlocking dimensions: (1) technological design and oversight through Dignity-by-Design (DbD), (2) limits to automation and meaningful human control, (3) contextual valuation, redistribution, and incentives, (4) political--institutional legitimacy and multi-actor governance, (5) sociocultural cohesion and the digital commons, and (6) legal--regulatory guarantees. The framework is operationalized through auditable organizational tools (principles, non-negotiable limits, and DbD checklists) aimed at aligning innovation with autonomy, equality, and human flourishing. Rather than offering a closed blueprint, we conclude by articulating open questions and tensions intended to foster interdisciplinary debate and guide future research.

\vspace{6pt}

\vspace{6pt}
\noindent\textit{All authors contributed equally to this work.}
\end{abstract}

\newpage

\section{Introduction}\label{sec:introduction}

No previous generation has experienced a technological transformation comparable in scale and scope to the one currently unfolding. A broad body of scholarship argues that contemporary digital transformation constitutes a historically singular reconfiguration of the material, economic, and institutional foundations of society \citep{Castells1996, Brynjolfsson2014}. We are living through what historians describe as ``critical junctures'' \citep{McChesney2007} or ``constitutive moments'' \citep{Starr2004}: relatively brief periods in which technological, political, and social change opens paths of development whose consequences are long-lasting and difficult to reverse. In the digital realm, this juncture is also marked by a transformation in the architecture of visibility: participation is increasingly mediated by systems that record, infer, and act upon behaviour continuously: an incipient Panopticon 2.0 in which being observable becomes a condition of ordinary social life.

From this perspective, Ferguson's \textit{The Square and the Tower} \citeyearpar{Ferguson2017} usefully frames technological disruption as a shift in the balance of power between hierarchies and networks. History, Ferguson argues, is shaped not only by formal, vertical institutions, but also by informal, horizontal networks that operate across and beneath them. Periods of hierarchical stability are repeatedly unsettled by agile networks that introduce innovation, dissent, or subversion, after which institutions attempt to regain control through regulation, incorporation, or co-optation. Contemporary digital platforms and social media are thus not an absolute novelty, but a technologically accelerated iteration of earlier communication revolutions, raising familiar questions about authority, legitimacy, and governance under conditions of heightened complexity.

At the end of the twentieth century, two major technological forces converged: the internet and mobility. This convergence gave rise to what is commonly known as Web 2.0 \citep{OReilly2005}, a phase in which individuals not only access a near-universal repository of information but also generate and disseminate content continuously and ubiquitously. This shift enabled the rise of digital platforms, redefining network effects and giving concrete expression to what \citet{Castells1996} described as the ``network society.''

Digital infrastructures now mediate core domains of social life: communication, commerce, work, education, health-related information practices, and political participation. These infrastructures are no longer peripheral tools. They operate as background conditions for social coordination and institutional functioning. Over the last three decades, networked computing, mobile connectivity, algorithmic processing, and data-intensive services have produced a shift that many scholars describe as structural rather than incremental: a reorganisation of how value is created, how coordination occurs, and how authority is exercised \citep{Castells1996, Brynjolfsson2014, vanDijck2014}. The transition from ``digitisation'' to ``datafication'' matters here. It marks a move from using digital systems to support pre-existing activities to using those systems to record, predict, and shape behaviour at scale, often through continuous data capture and algorithmic optimisation \citep{vanDijck2014, MayerSchonberger2013}. In this environment, platforms increasingly function not only as firms, but as intermediating infrastructures that organise markets, social relations, and public discourse through standard-setting, access control, and the design of interaction itself \citep{vanDijck2018, Bratton2016}.

As data collection becomes routine and inference becomes scalable, the practical terms of participation in the digital sphere are increasingly set by a small number of platforms. These terms are set in at least three ways. First, they are set through technical design: defaults, interoperability constraints, identity architectures, ranking and recommendation systems, interface frictions, and the systematic use of engagement metrics to optimise attention and behaviour \citep{vanDijck2018, Pasquale2015, Zuboff2019}. Second, they are set through contractual architecture: standardised terms, unilateral policy changes, and the combination of ``take-it-or-leave-it'' access conditions with complex privacy notices that few users can meaningfully evaluate \citep{Cohen2013, Solove2024}. Third, they are set through automated governance: large-scale content moderation, de facto speech rules, selective visibility, and algorithmic curation that determines what is amplified, suppressed, or monetised \citep{Gillespie2010, Pasquale2015, vanDijck2018}. This platform-based rulemaking increasingly operates alongside, and sometimes in tension with, the legal and moral ideals that support democratic societies. Rights such as autonomy, due process, equality, dignity and transparency may remain formally recognised, yet the everyday operational rules that shape digital life are often defined through private ordering and opaque systems rather than through public, contestable procedures \citep{Cohen2013, EDPS2015, Mantelero2022}. The result is a widening gap between the formality of rights and the lived experience of dependence on digital infrastructures for ordinary social and economic participation \citep{Zuboff2019, Veliz2020}. Seen through a panoptic lens, these arrangements approximate an architecture of asymmetric visibility: many are rendered legible and governable, while the mechanisms and purposes of observation remain largely opaque.

The panopticon metaphor captures several features of this shift. First, surveillance can be efficient: control scales when observation is automated and when users cannot verify when, how, or by whom they are being monitored. Second, the focus is not only on detecting deviance but on training and normalising behaviour, increasingly through statistical benchmarks derived from aggregated traces. Third, the locus of observation is dispersed: the ``tower'' is no longer a single vantage point but a distributed assemblage of devices, platforms, intermediaries, and models. Fourth, the result can be self-discipline and behavioural steering even without overt coercion, because interfaces, defaults, rankings, and recommendation systems make certain actions frictionless and others costly. Finally, technical illiteracy and structural opacity mean that many users cannot `see' the control room, even while it can see them.

The central problem is therefore not only privacy in the narrow sense of secrecy or confidentiality. It is the expansion of asymmetric informational power, a form of panoptic power that moves from episodic watching to continuous inference, classification, and behavioural shaping: the capacity to observe, infer, classify, and act on individuals and groups at scale, often across borders and with limited contestability \citep{Couldry2019, Susser2019}. This asymmetry grows more consequential---and more ethically problematic---as organisations shift from storing traces to producing profiles, predictions, and automated decisions that affect access to opportunities, visibility, and resources. Profiling and scoring systems, for example, can reproduce and amplify structural inequalities when the underlying data reflect existing social disparities and when models operate as opaque gatekeepers to credit, employment, housing, education, or reputational standing \citep{Barocas2016, ONeil2016, Mantelero2022}. These systems also enable forms of influence that are difficult to detect and contest in practice, including microtargeting, choice architecture manipulation, and other techniques of behavioural steering embedded in the attention-driven platform economy \citep{Tufekci2014, Susser2019, Zuboff2019}. In parallel, cross-context data aggregation and inference transform ``behavioural exhaust'' into sensitive attributes and vulnerability signals, intensifying the mismatch between what users think they disclose and what systems can infer \citep{Acquisti2016, MayerSchonberger2013}.

At the same time, data increasingly function as a production input and a source of durable market power. This is not merely because data support better prediction. It is because data interact with increasing returns to scale, complementarity with compute and talent, and network effects that can entrench incumbent advantages \citep{Jones2020, Varian2018}. Formal accounts of firm dynamics show how data accumulation can reinforce dominance through feedback loops: more users generate more data; more data improve products and targeting; improved products attract more users; and the cycle repeats \citep{Farboodi2019}. These dynamics complicate standard assumptions about contestability and entry, since advantages depend not only on capital investment, but on accumulated behavioural and relational data and on the learning processes built on top of them \citep{Farboodi2019, Jones2020}. As a result, competition concerns converge with governance concerns: market structure becomes tightly coupled to informational asymmetry, and both shape the practical room for user agency.

This challenges not only privacy but human dignity itself, insofar as it undermines autonomy and the status of individuals as co-authors of the social order \citep{deHingh2018}. Economically, the current data regime resembles a highly unequal and incomplete market: users supply an essential production input while capturing little of the resulting value, reinforcing proposals to conceptualise data as labour rather than as a costless by-product \citep{ArrietaIbarra2018}.

Regulatory frameworks, most prominently the GDPR and related EU initiatives, have strengthened baseline protections and clarified principles of lawful processing, purpose limitation, minimisation, and accountability \citep{GDPR2016}. Yet practical enforcement remains difficult in ecosystems characterised by cross-border operations, technical opacity, and rapid innovation cycles. Investigations and remedies often require audit access, technical expertise, and sustained capacity that many institutions struggle to maintain, especially when harms arise from complex chains of intermediaries and from probabilistic inference rather than from discrete, observable violations \citep{EPRS2021, Pasquale2015}. The result is a persistent mismatch between the formality of rights and the lived realities of platform dependence, where individuals may nominally hold protections but lack practical leverage, intelligibility, and effective exit options \citep{Solove2024, Cohen2013}.

A large interdisciplinary literature diagnoses parts of this mismatch, but it does so in partially separated strands. Political economy accounts explain how platform business models monetise behavioural data and attention, and how this can concentrate power and reshape both markets and the public sphere \citep{Zuboff2015, Srnicek2017, vanDijck2018, Couldry2019}. Legal scholarship documents the limits of notice-and-consent governance under conditions of complexity, lock-in, and information asymmetry, and questions whether individual choice can plausibly legitimise downstream data uses that are opaque, continuous, and inference-driven \citep{Bergemann2018, Solove2024}. Economic research clarifies why data differ from conventional goods, especially due to nonrivalry and scale effects, and why data accumulation can reinforce incumbency advantages, raising competition and welfare concerns that conventional tools often address only imperfectly \citep{Jones2020, Farboodi2019}. Finally, technical and policy work develops privacy-enhancing technologies and governance instruments, but it also shows that uptake is uneven where incentives favour maximising extraction rather than minimising risk, and where compliance is treated as a procedural layer rather than as a binding design constraint \citep{Acquisti2016, EDPS2015, HLEGAI2019}. Critiques from data studies and science and technology studies add that ``big data'' are never raw, neutral, or complete; they are shaped by selection, measurement, institutional incentives, and interpretive frames, which affects both accuracy and legitimacy in downstream decision-making \citep{Boyd2012, Gitelman2013, Kitchin2014}. Taken together, these strands converge on a shared diagnosis: contemporary digital participation is governed through infrastructures that generate structural asymmetries of power, while existing legal and institutional mechanisms often struggle to make rights effective under conditions of scale, opacity, and dependence.

In parallel, scholars and policy actors increasingly frame these tensions through the language of a ``digital social contract.'' The term appears in distinct traditions: workplace and organisational contexts shaped by digital visibility \citep{Berkelaar2014}, cybersecurity governance \citep{Clinton2016}, and normative proposals for a more just digital society \citep{Cardelli2020, Srinivasan2023}. Related debates on ``digital sovereignty'' have expanded rapidly, but they often diverge in focus: some emphasise state capacity and jurisdictional control, others stress industrial strategy and technological autonomy, and others prioritise user agency and rights \citep[e.g.,][]{Pohle2020, Hummel2021, Floridi2024}. These perspectives provide valuable insights, yet they rarely converge on a single, implementable framework that links normative commitments to operational governance across the full data lifecycle.

This paper addresses a specific gap that emerges from this fragmentation. Existing accounts often do one of four things. First, they offer powerful diagnoses, such as surveillance capitalism, data colonialism, platform oligopoly, but provide limited guidance on how to translate those diagnoses into enforceable design constraints and institutional duties. Second, they propose technical controls (including privacy-by-design and PETs) without a clear normative account of which uses should be prohibited even when technically feasible or nominally consented to. Third, they rely on individual consent and self-management as the primary legitimising mechanism, despite extensive evidence that individuals cannot realistically evaluate or negotiate the downstream consequences of modern data processing \citep{Solove2024}. Fourth, they treat sovereignty primarily as a state attribute, even though many of the practical levers that shape digital life operate through transnational private infrastructures and through collective-action problems that individual choice cannot solve.

Our contribution is to integrate these strands into a dignity-centric Digital Social Contract that is both normatively grounded and operationally oriented. The core move is conceptual and institutional. Conceptually, we defend a data-personalist position: personal data are not neutral inputs or ordinary commodities, but rights-laden emanations of the person that remain connected to autonomy, equality, and dignity \citep{Floridi2016, deHingh2018, EDPS2015}. This framing supports a rights-first baseline for personal data governance that does not depend on consent alone (notwithstanding the existence of exceptions or additional legal basis for the processing of personal data that in some jurisdictions can be applied for specific cases). Institutionally, we treat data sovereignty as a multidimensional construct that combines protection, participation, and provision, thereby connecting fundamental rights, legitimate governance, and practical guarantees that make those rights effective in real systems \citep{Abbas2024}. We also recognise that personal data increasingly operate as a production factor, which makes distributive questions unavoidable while not collapsing dignity into a price mechanism \citep{ArrietaIbarra2018, Jones2020}.

To make this framework usable, the paper develops three linked outputs that correspond to the gap identified above. First, Section~\ref{sec:framework} builds the theoretical foundations by connecting classic social-contract assumptions to the digital condition and by clarifying why inference-driven systems and cross-border platform infrastructures strain territorial enforcement, democratic legitimacy, and meaningful agency. Second, Section~\ref{sec:dimensions} translates the normative stance into six operational dimensions for dignity-centric governance, technological, ethical, economic, political-institutional, sociocultural, and legal-regulatory, so that governance is not reduced to a single discipline or instrument. This section also proposes Dignity-by-Design (DbD) as a practical method for linking human-rights commitments to system requirements, auditability, and enforceable organisational duties. Third, Annex~1 proposes a research and evaluation agenda designed to support cumulative progress rather than one-off prescriptions, including interdisciplinary metrics and stress tests that can be applied across sectors and jurisdictions.

This approach differs from prior ``digital social contract'' proposals by combining four elements in a single framework: (i) a dignity-based account of the moral and legal status of personal data; (ii) a sovereignty model that integrates rights, agency, and operational capacity rather than treating control as an individual interface feature; (iii) an explicit bridge from normative claims to design and governance obligations through DbD; and (iv) an evaluation programme oriented toward implementation, measurement, and institutional learning. At the same time, our proposal remains continuous with earlier work: it adopts the social-contract lens to organise legitimacy questions, draws on established analyses of platform power, and treats technical governance as necessary but insufficient unless supported by enforceable institutional arrangements.

Relative to the Global Digital Compact and the Council of Europe AI Convention, this work positions itself at the ``meso-level'' between principle-setting and implementation. The Global Digital Compact provides a broad agenda for digital cooperation, covering a safe and secure digital space, data governance, and international AI governance, but, by design, it remains general and programme-like. The Council of Europe's AI Convention offers a rights-based lifecycle frame for AI activities, yet it does not on its own resolve the wider political economy of the data economy (inference markets, platform lock-in, and the distribution of data-generated value). The dignity-centric Digital Social Contract advanced here is therefore complementary: it identifies the inference layer as the primary locus of contemporary power, proposes categorical red lines where nominal consent cannot legitimise extraction or high-stakes automation, and offers a concrete design-and-audit method (Dignity-by-Design) to translate human-rights commitments into system requirements that organisations and regulators can test. In this sense, the framework can function as a practical bridge for implementing GDC commitments and operationalising the AI Convention's lifecycle governance, while remaining consistent with their shared human-rights orientation and multi-actor governance ambitions.

The remainder of this paper is organized as follows. Section~\ref{sec:framework} develops the theoretical foundations of a data-personalist social contract. Section~\ref{sec:dimensions} articulates six operational dimensions for dignity-centric personal data governance. Section~\ref{sec:discussion} proposes a research and evaluation agenda. For reasons of length, the operational policy blueprint that translates the six dimensions into implementable institutional levers and enforcement principles is provided in Annex~1. To enhance conceptual clarity and avoid terminological ambiguity, Annex~2 sets out explicit definitions of the core concepts used throughout the paper, specifying the precise meanings and theoretical assumptions adopted in our framework.

\section{Theoretical Framework: From Classic Social Contracts to a Dignity-Centric Digital Social Contract}\label{sec:framework}

\begin{figure}[htbp]
\centering
\includegraphics[width=0.9\textwidth]{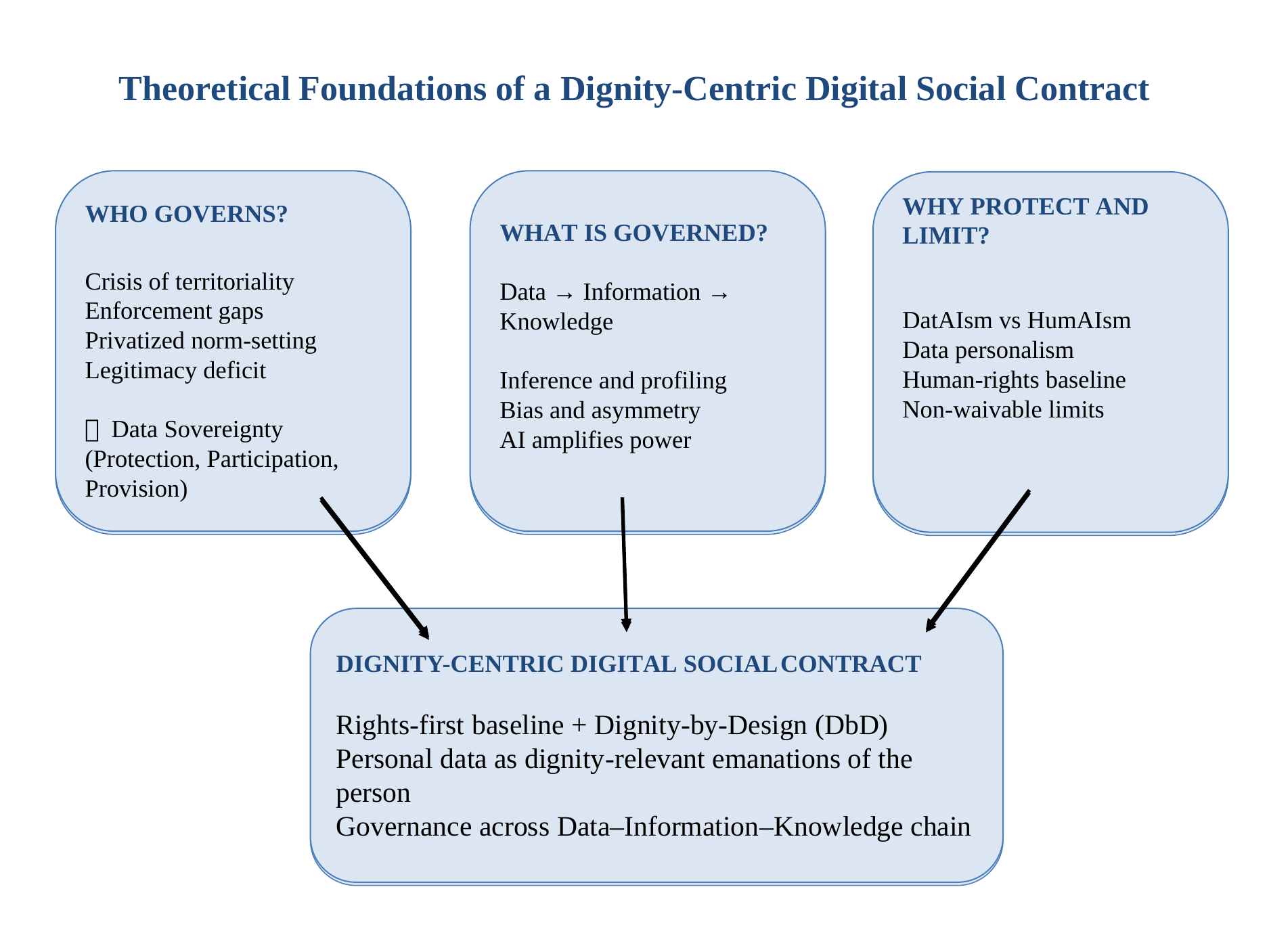}
\caption{Foundations of a Dignity-Centric Digital Social Contract.}
\label{fig:foundations}
\end{figure}

We build the theoretical framework for a dignity-centric Digital Social Contract in three main steps (Figure~\ref{fig:foundations}). First, under the question \textit{who governs} we recall the classic social contract and show how digital infrastructures strain its territorial, juridical, and subjective foundations through a connected sequence of failures in territoriality, enforcement, norm-setting, and legitimacy. Then, we reframe the response in terms of data sovereignty, understood as protection, participation, and provision. We move then to the question \textit{what is governed} and clarify how data are transformed into information and knowledge, and why inference and bias turn epistemic processes into forms of power---an issue that AI amplifies but does not create ex nihilo, since many AI privacy failures are best understood as remixed and intensified versions of longstanding data-collection, inference, and decision-making problems \citep{Solove2025}. Finally, we focus on the question \textit{why protecting and limiting}, in which we emphasize why protection requires institutions. Then we contrast DatAIsm and HumAIsm and introduce data personalism as a normative position about the human person. Afterwards, we ground personal data in a human-rights baseline that cannot rely on consent alone or as the main lawful basis and recognize personal data as a production factor, to draw the corresponding distributive implications. Finally, we propose a Dignity-Centric Digital Social Contract as the institutional solution to these failures: a framework that re-anchors data governance in human dignity, makes protection enforceable across the full data--information--knowledge chain, and coordinates rights, design obligations, and distributive mechanisms so that innovation remains possible without reproducing extraction, opacity, and asymmetric control. This framework then motivates the dimensions and design principles developed in Section~\ref{sec:dimensions}.

\subsection{Who governs}\label{sec:whogoverns}

\subsubsection{Classic Social Contracts and the Digital Condition}\label{sec:classic}

Although the origins of the social contract theory can be found in Plato and Socrates, Hobbes, Locke, and Rousseau popularized the concept in the eighteenth century \citep{Hobbes2012, Locke1988, Rousseau1997}. More recently, led by John Rawls, a great variety of social contract theories have been proposed \citep{Liaropoulos2020}. All versions attempt to explain why rational individuals would voluntarily consent to give up their natural freedom to form and legitimate a society where living together under common laws and social enforcement mechanisms.

The classical social and political architecture, which would permit people collectively enjoy the benefits of political order, is based on three components: (a) national identity: citizen as a subject of a state; (b) legal or juridical identity: citizen as a holder of rights and duties and (c) subjective identity: citizen as an aware, rational and affective being \citep{Boucher1994}.

The evolution of the XXI century is showing some paradigmatic contradictions with the social contract described in each of its three elements.

The digital condition stresses the classic juridical architecture of the social contract through a connected sequence of failures: territoriality, enforcement, normative displacement, and legitimacy.

First, there is a \textit{crisis of territoriality}. Digital life is organised through transnational infrastructures (platforms, clouds, content distribution, cross-border data flows) in which actors, processing operations and effects are distributed and often opaque. This complicates the basic public-law coordinates that make rule-making and adjudication workable: identifying the competent authority, determining the applicable law, and allocating responsibility along technically fragmented chains.

Second, this deterritorialisation translates into a \textit{crisis of enforcement}. Even where legal systems claim extraterritorial reach (through market- or effects-based connecting factors), effective enforcement depends on investigatory capacity, access to evidence, auditability of socio-technical systems, and credible coercion against globally scaled firms. The result is a structural gap between formal validity and practical effectiveness: rights may exist in law, yet remain difficult to exercise and even harder to vindicate where harms arise from opaque profiling, inference, and automated decision-making.

Third, the enforcement gap accelerates a \textit{privatisation of normativity}. In practice, a significant portion of the rules that govern participation in the digital sphere are produced and executed through private ordering and technical architecture: \textit{terms and conditions}, unilateral policy updates, contractual standardisation, interface design, recommender logics, app-store governance, and access controls. This is not merely contractual freedom; it is a form of infrastructural regulation in which compliance is achieved by design (through defaults, friction, ranking, exclusion or deplatforming) rather than through the guarantees and publicity typical of public law.

Finally, this displacement produces a \textit{legitimacy deficit}. When de facto rules shape rights, opportunities, and public discourse while being set by private actors and enforced through automated systems, the classical sources of normative authority weaken: democratic authorship, transparency, due process, proportionality, reason-giving, and effective remedy. The core problem is therefore not only regulatory capacity, but justification: who sets the rules of digital life, under what title, with which safeguards, and with what accountability to those subject to them.

This four-step sequence provides the legal background for the tensions analysed below in the three classic components of the social contract: territorial identity, juridical identity, and subjective identity.

\paragraph{A) Territoriality.}

The state-nation is firstly a territory with well delimited borders that generates specific sociopolitical realities, which bind the citizens of that state-nation but not foreign people. In the same way, the direct effects of those sociopolitical realities are circumscribed to the interior of territory. Thus, territoriality makes it possible to determine who is in and who is out of these realities, who is a citizen and who is not. Only citizens subscribe to the social contract.

Globalization, digitalization and technological disruption, which characterize the Fourth Industrial Revolution, are breaking the stable territoriality framework described above. Although the revolution is based on a technological upheaval, its most important effect of this is its socio-political nature \citep{Chandler2015}.

(a) One of the effects of the integration of many more countries into the global economy is the blurring of the state's frontiers and the change of their governance structures \citep{Wahl2014}, the deterritorialization of politics \citep{Choucri2012}. The widespread globalization impact contrasts with the absence of any overarching global authority capable of facilitating competitive global markets, undermining the effectiveness of regulation and bypassing mechanisms of responsibility \citep{Spraul2021, Scherer2016}.

(b) The second factor which explains the death of the territoriality as main classical social contract element is the cyberspace, a territory with ubiquitous nature, vast scale and scope, instantaneity, and nontransparent and often complex interconnections. This new reality reshapes contemporary theory, policy, and practice. While with globalization territoriality has not ceased to exist, but has extended to blur borders, cyberspace extends borders so much that it renders them useless, because in a global and hyper-connective society, states can no longer be the sole security providers \citep{Clinton2016} requiring the coordination of all stakeholders. Cyberspace constitutes a fundamentally new domain that, together with the users' evolving habits, requires fundamentally new government system or new international legal instruments.

(c) While globalization weakens and cyberspace makes useless state sovereignty by extension, disruptive technology (the triumvirate of data/algorithms/AI) destroys territoriality in its roots. Following Hobbes, we can distinguish between (1) communities established by institution, through a mutual agreement of free individuals, and (2) communities formed by acquisition, by conquering a pre-existing sovereignty. In the former, subjects authorize and make the sovereign their representative. This authorization requires a real contract: the subject obeys the sovereign because they have agreed to do so and, moreover, because they have agreed to authorize the punishment of those who disobey. Conversely, in communities formed by acquisition, foreigners, now citizens, are forced to accept the new sovereignty and relinquish the previous one without authorization, but with an obligation of obedience and all the associated responsibilities. The triumvirate transforms people into technological citizens subjected to a cyberspace of vast capabilities without their conscious consent. Within it, they are responsible, yet the state to which they belong cannot provide them with the necessary security. \citet{Shin2025} argue that algorithmic fact-checking constitutes much more than a technical intervention; it marks a paradigmatic shift in how truth is operationalized, distributed, and institutionalized in the digital age. Because of this, these AI-based systems demand not only technical scrutiny but philosophical and sociotechnical reflection, one that situates algorithmic truth within broader infrastructures of knowledge governance.

The asymmetry of power between citizens and those powerful \textit{outsiders} creates vulnerabilities that threaten democratic values and human rights. Speaking about big data, \citet{Couldry2019} argue that this reflects a new form of colonialism, where global platforms extract value from populations in ways that replicate historical patterns of exploitation.

\paragraph{B) Legal or juridical identity.}

Since legal rights are morally plausible as products of legitimate institutional processes, states defend and uphold them. Institutional development, if it occurs, must at least justify the same procedural legitimacy. However, when territoriality erodes, impacting the intrinsic values of participation, transparency, neutrality, and fairness in procedural justice \citep{Yang2024}, at least two basic problems arise: (a) the emergence of open disputes over the governance of the new space, and (b) the emergence of stakeholders who, being external to governance, have the power to impose rules.

(a) To the extent that states cannot guarantee the social contract of a global and hyperconnected society, they are forced to share their legitimacy with international and non-governmental institutions, private companies, and other non-state actors. For example, while the effectiveness of global regulatory control is declining, proactive and flexible corporate action is increasing, with voluntary self-regulation being a strategic response to the new environment \citep{Vitiea2019}. At the same time, a wide range of actors and groups, and governance systems that did not exist a few years ago, are emerging, with which the State is now forced to share its authority \citep{Qiao2022}. The central question, undoubtedly, is which norms, regulations, and institutions legitimately ``close'' the structure of this new space. Several authors who describe the end of governmental primacy and the evolution of public regulation point to the need for a polycentric governance model with new forms of multilateral and non-territorial regulation \citep{Backstrand2017}.

(b) When Mark Zuckerberg \citeyearpar{Zuckerberg2017} described Facebook as a ``Global Community'' that would govern the planet in most aspects of life, where each user could vote on ``global issues that transcend national borders'' and ``participate in collective decision-making,'' he was thinking of new rules unrelated to those described in the previous point. That is, of new actors and new rules not legitimized through established channels, in what some have called a silent transfer of sovereignty to third parties \citep{Cassese2003}. We refer to the emergence of large and powerful outsiders, citizens of nowhere, such as platform monopolies as Google, Facebook, Apple and Amazon, which, although they do not possess physical territory, can deeply influence any citizens \citep{Shadmy2019} without citizens being protected by the law and regulations, the police, the courts or other similar institutions typical of the social contract. In this sense, we agree with \citeauthor{Bratton2016}'s (\citeyear{Bratton2016}) idea: platforms are not technical architectures but institutional structures. The ability of the states and markets to regulate platforms is low \citep{Shadmy2019}, being undermined due to monopolistic circumstances \citep{AlRodhan2014}. New regulations and structures are needed to ensure that the contract of adhesion will not be used solely for commercial or partisan purposes, but rather for broader societal goals and digital public good \citep{Ghosh2018}. Such laws must achieve two main things: (a) they must restrain those entities which control a disproportionate amount of data, and (b) they must ensure the enforceability of such laws.

\paragraph{C) Subjective identity.}

The social contract requires a voluntary agreement between individuals, resting upon the trust on a specified, undertaken, fulfilled, and fairness system. If potential signatories were to doubt, it would no longer be rational for them to contract or cooperate. States should allow for legal security that enables citizens to perceive satisfaction and tranquility by observing how the catalog of values possessed by the legal system, which is the product of their agreement, is guaranteed and, in turn, how it materialized. Moreover, social contract presupposed something like a rough equality of stakes.

The individual is rationally oriented towards the achievement of their basic interests within a regime of shared sovereignty that unifies them and allows them to freely and voluntarily subordinate their particular interests to the larger whole. In this system, individuals develop a sense of shared membership oriented toward what unites them in a common political order, rather than what divides them or is imposed upon them \citep{Rousseau1997}. These sovereign individuals share a similar model of conduct, way of life, or expectations about how others will behave, so that they interact with each other in terms of social and familial contractual relationships no less than in their political life.

Drawing on a social contract perspective, \citet{Abbas2024} distinguish three interrelated facets of sovereignty: (a) protection (upholding fundamental rights and safeguards), (b) participation (enabling meaningful individual and collective agency in governance), and (c) provision (deploying technical, organizational, and legal-operational guarantees that make those rights real).

In the triumvirate conditions that leaves a public quite at the mercy of the outsiders, which controlling vast reserves of personal data possesses predictive powers enabling it to have an unequal stake in the shared system, and fortify the possessor against any form of reprisal, cooperation could not be rational. At the limit, \citet{Bratton2016} suggests seeing AI systems as part of a vast, layered apparatus of control and coordination, where epistemology is no longer tethered to human subjectivity.

Thus, a ``social contract 2.0'' requires reimagining the conditions under which epistemic legitimacy and democratic control are constructed in datafied environments. It requires that the classical elements of contractarian legitimacy---trust, reciprocity, and a rough equality of stakes---be made institutionally credible under contemporary informational asymmetries. The dominant approaches in this field fall into three main categories \citep{Pohle2020}:

\begin{enumerate}[label=(\alph*)]
\item those that advocate for state control over digital infrastructures and the capacity to design and implement digital policies, often with an emphasis on cybersecurity;
\item those that advocate for the control of technological monopolies by creating national technology companies and developing effective industrial and innovation policies; and
\item those that advocate for the right to digital self-determination, user agency, and the capacity to make informed decisions about personal data and algorithmic environments.
\end{enumerate}

We believe we need a holistic approach, one that is not based only on what states can do as primary actors (since the erosion of territoriality is very significant) but rather on the development of extraterritorial systems based on what unites us: individual data sovereignty.

\subsubsection{Individual Data Sovereignty as a Multidimensional Construct}\label{sec:sovereignty}

As mentioned, from a social contract perspective, data sovereignty is best understood not as mere ``control'' but as a multidimensional construct that integrates \citep{Abbas2024}:

\begin{itemize}
\item Protection: rights and safeguards such as privacy, security, and non-discrimination;
\item Participation: roles and responsibilities of actors in data governance;
\item Provision: operational guarantees---control, compliance, and safe sharing across contexts.
\end{itemize}

From a welfare-economics perspective, the protection--participation--provision triad can be read as an institutional map of how societies constrain, legitimate, and implement data governance. Protection defines the feasible set by placing non-waivable limits on data practices that would undermine dignity, autonomy, or equality. Participation specifies whose interests count and how individuals and groups can exercise agency over the rules and institutions that govern data environments, addressing collective-action problems that individual choice cannot solve. Provision supplies the shared capacities (standards, audits, enforcement, and usable controls) that make rights credible rather than merely formal. This mapping clarifies why data sovereignty is not reducible to ``individual control'' over isolated records: it is a social-choice problem about the governance of an informational environment with distributive consequences and externalities.

Contemporary debates on ``digital sovereignty'' highlight how this construct is claimed and interpreted by different actors. \citet{Couture2019} show that ``sovereignty'' in the digital field is invoked to assert various forms of collective control over infrastructures, technologies and data. \citet{Fratini2024} map four main models of digital sovereignty---rights based, market oriented, centralisation, and state based---none of which fully reconciles effective regulation with technological dynamism. In parallel, the edited volume \textit{Data Sovereignty: From the Digital Silk Road to the Return of the State} \citep{Chander2023} examines how states seek to reassert territorial control over crossborder data flows.

A crucial strand of this debate concerns collective claims over data. The literature on Indigenous data sovereignty, exemplified by \citeauthor{Kukutai2016}'s (\citeyear{Kukutai2016}) foundational volume, argues that communities have inherent rights to govern data about their peoples, lands and cultures, as part of broader self-determination. This perspective resists framing data merely as an individual property right and emphasises relational, historical and political dimensions.

Our guiding claim is that a dignity-centric digital social contract must integrate these layers. It must:

\begin{enumerate}
\item Uphold fundamental rights (Personal Data protection and privacy, nondiscrimination, due process, informational self-determination), not only in law but also in technical architectures.
\item Enable meaningful agency---both individual and collective---over how data are generated, processed, and valorised.
\item Deploy governance mechanisms---technical, organizational and legal---that make those rights real, while addressing economic justice in data markets, including questions of benefit sharing and fair compensation.
\end{enumerate}

In this sense, data sovereignty is neither a purely state-centric nor a purely individualistic concept. It is a negotiated arrangement among persons, communities, firms, and states about who has which rights and duties over the life cycle of data, under what safeguards, and for whose benefit.

To make this multidimensional sovereignty operational, we must also clarify what ``data'' are in practice,---how they become information and knowledge, and where normative stakes intensify along that transformation as the object of government.

\subsection{What is governed}\label{sec:whatgoverned}

In the current context, effective data governance requires first clarifying what, exactly, is being governed: what counts as data, how data are generated, and under which assumptions data can support knowledge claims. Analysing the metaphysics and epistemology of data is therefore crucial, because it makes explicit the conceptual, methodological, and normative choices embedded in data collection, modelling, and interpretation, helping to identify sources of bias, uncertainty, and error. It also brings into view the underlying metaphysical assumptions about what data are taken to be---representations, abstractions, or mediated phenomena---which shape what kinds of claims data can legitimately support.

The classic data--information--knowledge--wisdom (DIKW) ladder remains a useful starting point for analyzing the metaphysics and epistemology of data, as originally articulated by \citet{Ackoff1989} from Information Theory:

\begin{itemize}
\item Data: symbolic representations of states of the world; traces that may or may not be personal and that, taken alone, have relatively low interpretive content.
\item Information: structured or aggregated data that reveal relationships or meaning (e.g., profiles), enabling inferences about habits, preferences, or intentions.
\item Knowledge: information integrated into a context and embodied in an agent with an intention of use, enabling decisions and action.
\end{itemize}

However, its apparent simplicity conceals a wide variety of interpretations that depend on the disciplinary lens through which the ladder is approached. Information theory \citep{Shannon1948}, economics, law, computer science, philosophy of science and the social sciences all mobilise the DIKW framework differently, emphasising distinct transitions, risks and normative implications along the data value chain. Rather than a single linear or neutral progression, the ladder functions as a conceptual space in which epistemic, economic, legal and political assumptions are made visible.

From the standpoint of economics and finance, this ladder is interpreted primarily as a value-generation chain \citep{CastroFernandez2025}. Data are understood as a strategic, non-physical asset whose economic value does not reside in raw traces as such, but emerges progressively as data are transformed into information, knowledge and, ultimately, actionable decisions. Within this framework, established valuation paradigms---cost-based, market-based and use-based approaches---are mobilised to estimate the contribution of data to firm value. Replacement costs, observable prices in emerging data markets, and the marginal impact of data on predictive models, key performance indicators and future revenues all serve as proxies for this contribution. This perspective also highlights how processes of ``data assetization'' underpin contemporary platform business models, in which user behaviour and attention metrics function as core intangible assets, and why ongoing debates on regulation and data governance are central to stabilising credible methods for pricing and accounting for data in the economy. A complementary strand in economic sociology and Science and Technology Studies (STS) emphasizes that ``data value'' is not a pre-given property of datasets but the outcome of assetization: measurement practices and governance devices that define boundaries and translate uncertain future gains into capitalized claims. Empirically, in platform markets this often occurs through the assetization of users and engagement metrics (e.g., DAU/MAU, ARPU) rather than through explicit balance-sheet recognition of personal data as an owned asset. Accounting regimes can further limit the recognition of internally generated data as separable intangible assets, so financial markets may price data-intensive business models indirectly through market capitalization and goodwill \citep{Muniesa2017, Birch2020, Birch2021, Nani2023}.

From a technical and privacy-engineering perspective, the DIKW ladder reveals a gradient of risk and power asymmetry. As data move upward along the ladder, the potential for harm increases: raw data typically entail low to medium risks; information enables profiling and thus medium to high risks; and knowledge embedded in automated or semi-automated decision systems generates high-risk asymmetries of power, particularly under conditions of large-scale automation. While privacy-enhancing technologies and good system design can mitigate these risks, they cannot eliminate the structural amplification of power that accompanies higher-level inference and decision-making.

From the perspective of data science and the social sciences, knowledge production along the DIKW ladder is neither linear nor neutral. Data do not enter the chain as raw, objective inputs: they are selected, captured and formatted according to prior assumptions, institutional incentives and specific purposes, which already shapes what can later count as information. In turn, information is not simply ``revealed'' by processing data, but constructed through modelling choices, classifications and interpretive frameworks that embed social and power relations into apparently technical outputs. Finally, when information is stabilised into knowledge---through metrics, predictive systems and authoritative narratives---it feeds back recursively into the world, influencing what is subsequently observed, recorded and treated as relevant. This undermines the common idea of Big Data as an exhaustive or theory-free representation of reality, and instead highlights that the DIKW chain is always partial and contestable, raising questions about whose data are collected, for what ends, and who can access, interpret or challenge the resulting knowledge. Importantly, these epistemic dynamics also have sociological consequences: in data-driven digital platforms, the very extraction of data from everyday life often relies on a continuous erosion of privacy, normalising the commodification of intimate behaviours and social ties in ways that can corrode ordinary trust and conditions for self-development \citep{Veliz2020}. In this sense, the DIKW ladder is not only a model of knowledge production but also a mechanism through which social relations, dignity and vulnerability---particularly among adolescents---can be reorganised and governed via datafied environments \citep{Floridi2016}.

From a legal and ethical standpoint, each transition along the data--information--knowledge chain introduces additional normative constraints. Knowledge is not a simple aggregation of information, but a fallible and bias-prone output of statistical analysis, inference and machine learning. As such, it demands robust guarantees of quality, traceability and explainability. This perspective also foregrounds the structural asymmetries of contemporary data markets, in which individuals typically lack informed consent, bargaining power and effective contractual protection. Within this framework, data protection law, AI regulation and governance mechanisms are interpreted as corrective tools aimed at addressing these failures in the production, circulation and monetisation of data, information and knowledge.

AI intensifies the governance relevance of the DIKW ladder because it systematically converts traces into inferred attributes, scores, and predictions---i.e., it industrializes ``data generation'' at the knowledge stage. When organizations can create new personal data via inference, restrictions that focus only on what is ``collected'' risk becoming partially illusory: inferred data can subvert public expectations and enable functional end-runs around legal and ethical constraints designed for direct collection. For this reason, inferred data should be treated on an equal footing with collected data for purposes of constraint, contestability, and remedy; otherwise, the most power-laden outputs of the data value chain remain structurally under-governed \citep{Solove2025}.

The economic logic also clarifies why governance must target the inference layer, not only the collection of ``raw traces.'' In DIKW terms, most private rents are captured at the information and knowledge stages (profiling, prediction, ranking, and decision), where downstream uses become consequential and where bargaining-power asymmetries and externalities intensify. A rights-first contract therefore cannot treat downstream inference as a technical by-product: as processing moves from traces to profiles and consequential inferences, purpose proportionality, traceability, and impact assessment become central economic safeguards rather than mere compliance overhead \citep{Goldfarb2011, ArrietaIbarra2018, Frey2024}. Additionally, economic analysis further extends this view by framing informational risks and harms in terms of informational externalities. From this angle, operations that degrade data quality, increase opacity or amplify bias generate social costs that are not adequately internalised by current market structures. A Pigouvian approach \citep{Pigou1920} to information governance would therefore require such externalities to be measured, priced and internalised, rather than treated as unavoidable side effects of data processing. Making these costs visible---conceptually and institutionally---would guide both regulation and system design in a manner analogous to how Pigouvian taxes function in environmental policy \citep{GarridoMerchan2026}.

Philosophical analysis deepens these concerns by distinguishing between different metaphysical conceptions of data. One view treats data as representations: syntactic data, semantic information and pragmatic knowledge understood as signs pointing to factual states of the world. A second view, prevalent in machine learning, treats data as abstractions, reducing entities to measurable attributes while bracketing questions of meaning and truth as irrelevant to engineering performance---an approach that enables large-scale modelling but risks subordinating reason to mere calculability. A third view, rooted in the philosophy of science, emphasises that we never access ``things in themselves'' but only phenomena: what counts as a datum is always mediated by prior concepts, purposes and standards of relevance. On this account, so-called ``raw data'' are an oxymoron; data are always already ``cooked.''

Taken together, these perspectives underscore that digital social and cultural data consist of undecoded traces of social life, generated and filtered by infrastructures and business models whose logics remain largely opaque to researchers and citizens alike. Methodological bias---fragmentary, selective and decontextualised traces---is therefore not a marginal flaw but a structural feature of datafied societies. Analyses of algorithmic systems as ``weapons of math destruction'' illustrate how biased data and opaque models can entrench inequality and undermine democratic accountability.

Beyond epistemic concerns, this asymmetry in data and inference constitutes a form of power. Control over personal data translates into the capacity to shape preferences, discriminate between individuals and groups, and erode the conditions of democratic deliberation. From this perspective, privacy is not merely an individual preference or a tradable commodity, but a precondition for autonomous agency and collective self-government. This insight aligns with the datapersonalist stance developed in this project: treating personal data as right-laden manifestations of the person, rather than neutral inputs, is both an epistemic and a political necessity.

Finally, from an engineering and governance perspective, data sovereignty emerges not only as a legal or institutional problem but also as a technical one. Privacy by design and accountability by default become concrete engineering obligations oriented toward the protection of human dignity. Crucially, this approach demonstrates that technological design can be explicitly integrated with social-science and normative perspectives, rather than opposed to them, and that such integration is feasible in practice. Accordingly, we believe that the supposed tension between innovation and rights constitutes a false dilemma: far from being mutually exclusive, technological innovation and the robust protection of fundamental rights are not only compatible but mutually reinforcing when data governance is grounded in human dignity.

Once data are understood as part of an epistemic and economic pipeline rather than neutral ``facts'', and thus, the object of the question \textit{what is governed}, the next question to be analysed is \textit{why protecting and limiting} personal data harvesting and use along the DIKW chain.

\subsection{Why protecting and limiting personal data}\label{sec:whyprotecting}

The need to govern the use of and protect personal data requires a structural response to how contemporary digital systems convert traces into profiles, predictions, and consequential decisions across the data--information--knowledge chain. These transformations redistribute power, shape life chances, and can undermine the conditions for autonomy, dignity, and democratic legitimacy. For this reason, the ``why'' of protection and limit cannot be reduced to individual consent or technical risk management alone or as the main legal basis: it must be justified at the level of social institutions, normative anthropology, and political economy. We proceed in four steps. First, we provide a critical genealogy showing why ungoverned technological acceleration predictably concentrates power and value, making institutional countermeasures a precondition for credible protection. Second, we contrast competing anthropologies of data---DatAIsm, HumAIsm, and data personalism---to clarify which conception of the person can legitimate limits on data extraction and inference. The terms DatAIsm and HumAIsm are introduced here as analytical ideal-types rather than as established categories in the literature. They synthesise identifiable intellectual currents---data-driven positivism and human-centered normative traditions---into heuristic constructs that clarify the underlying anthropological assumptions shaping contemporary digital governance debates. Third, we translate this datapersonalist stance into a rights-first baseline that cannot rely on consent alone or as the main legal basis and that targets harms emerging at the information and knowledge stages. Finally, we recognise personal data as a production factor to draw the corresponding distributive implications, clarifying why benefit-sharing and incentives must be addressed within---rather than in place of---a non-waivable dignity-based floor.

\subsubsection{From the General Intellect to Platform Capitalism: A Critical Genealogy}\label{sec:genealogy}

Analytical tradition has diagnosed the structural risks of ungoverned technological acceleration with remarkable prescience. Marx, in the Grundrisse \citeyearpar{Marx1973}, anticipated that the development of productive forces would reach a stage at which direct human labour ceases to be the principal source of wealth, replaced instead by the ``general intellect'' embodied in machinery. This observation, read as a descriptive claim rather than a normative programme, captures with notable accuracy the current dynamics of automation and artificial intelligence.

The Frankfurt School extended this analysis to the cultural and political dimensions of technological domination. \citet{Marcuse1964}, in \textit{One-Dimensional Man}, argued that advanced industrial society integrates potential opposition by channeling technological rationality into systems of control that flatten critical thought and reduce individuals to functional elements of the productive apparatus. This insight anticipates the contemporary concern that algorithmic systems may not merely automate labour but also shape preferences, constrain choices, and erode the conditions for autonomous deliberation.

\citet{Negri1991}, in \textit{Marx Beyond Marx}, re-read the Grundrisse to argue that the ``general intellect'' was not merely an objective force embedded in machines but a living capacity of collective labour, whose value is systematically captured by capital. \citet{Hardt2000} developed this thesis in \textit{Empire} and \textit{Multitude} \citep{Hardt2004}, contending that in the era of ``immaterial labour'', characterised by the production of information, communication, and affect, the boundaries between production and life dissolve, enabling an unprecedented extraction of value from social cooperation itself. \citet{MoulierBoutang2011} coined the term ``cognitive capitalism'' to describe this regime, in which the primary source of surplus value shifts from physical labour to the cognitive, relational, and affective capacities of networked individuals.

\citet{Deleuze1972}, in \textit{Anti-Oedipus}, had already envisioned that the progressive replacement of living labour by machines would generate a structural crisis of employment whose resolution demands rethinking the distribution of social value. \citet{Cammatte1973} warned that capital has autonomised itself from human will, so that any intervention failing to alter the internal logic of accumulation is condemned to futility. These diagnoses were updated and synthesised by \citet{Williams2013} in the \#ACCELERATE MANIFESTO, which argued that technological progress should not be resisted but redirected through democratic governance, so that automation serves as a material condition for collective well-being rather than precarisation. \citet{Srnicek2015} elaborated this argument in \textit{Inventing the Future}, proposing the reduction of the working week, the implementation of universal basic income, and the democratic control of productive technologies as institutional responses to automation. \citet{Srnicek2017}, in \textit{Platform Capitalism}, further showed how digital platforms extract value precisely by positioning themselves as indispensable intermediaries that capture the data generated by user activity, a mechanism that recasts classical rent theory in informational terms.

\citet{DyerWitheford2015}, in \textit{Cyber-Proletariat}, extended this genealogy by documenting how digital networks simultaneously create new circuits of global exploitation and new forms of precarious digital labour, from content moderation in the Global South to gig economy logistics. \citet{Mason2015}, in \textit{PostCapitalism}, argued that the informational character of contemporary production, the near-zero marginal cost of reproducing digital goods, undermines the price mechanism itself and points toward the necessity of post-market forms of coordination. \citet{Bastani2019}, in \textit{Fully Automated Luxury Communism}, pushed the argument further, contending that automation, renewable energy, and synthetic biology could, under appropriate institutional conditions, eliminate scarcity for essential goods.

Independently of the intellectual tradition from which they originate, these analyses converge on a single essential point: without institutional mechanisms for redistribution and democratic governance, and thus, mechanism to protect and limit the harvesting and use of data, technological acceleration concentrates power and wealth rather than diffusing them.

\subsubsection{Competing Anthropologies of Data: DatAIsm, HumAIsm, and Data Personalism}\label{sec:anthropologies}

The preceding sections have shown that data sovereignty and data governance cannot be adequately understood without engaging the epistemic and metaphysical assumptions embedded in data-driven systems. However, these assumptions are themselves grounded in deeper philosophical paradigms that articulate competing visions of the human person, knowledge, and social order in the digital age. Revisiting these paradigms is therefore essential to assess which forms of data governance are compatible with human dignity, democratic legitimacy, and meaningful agency. This section examines three influential and contrasting frameworks---DatAIsm, HumAIsm, and Data Personalism---not merely as abstract philosophies, but as operative worldviews that already shape digital infrastructures, business models, and regulatory imaginaries. By making these paradigms explicit, the analysis clarifies why incremental regulatory fixes are insufficient and why a new digital social contract is required to re-anchor technological development in a defensible conception of the human person.

DatAIsm is a positivist, utilitarian and nihilistic philosophy that reduces humans from subjects of action and meaning to observable behaviour, and from citizens to consumers. It assumes that everything that matters can and should be reduced to data; that large datasets yield superior, quasi-absolute knowledge; and that correlation-driven prediction can substitute causal explanation and reflective judgment. In practice, DatAIsm underpins business models that treat human experience as ``raw material'' for behavioural prediction markets, echoing \citeauthor{Zuboff2019}'s (\citeyear{Zuboff2019}) account of \textit{surveillance capitalism}, in which behavioural surplus is extracted and traded without meaningful consent or reciprocity.

Historically, DatAIsm does not emerge in a vacuum but can be seen as the latest expression of longer intellectual trends: positivism's faith in observable facts, utilitarianism's drive to compress value into a single maximisable metric, behaviourism's focus on observable responses rather than inner states, and contemporary tech solutionism, which treats complex social and moral problems as optimisation tasks to be solved by more data and more computation. This genealogy helps explain why DatAIsm finds such a natural home in data-driven platform capitalism and why its assumptions so often go unchallenged.

This logic is intertwined with new critiques of the digital political economy. \citet{Varoufakis2023} argues that dominant platforms increasingly resemble ``technofeudal'' lords: they control key digital infrastructures, extract rents (``cloud rents'') rather than competing in open markets, and reduce users and complementors to digital vassals. DatAIsm provides the epistemic and ideological justification for this order: if data driven optimisation is the overriding goal, then centralised surveillance and behavioural nudging become not only permissible but desirable.

HumAIsm, by contrast, defends the primacy of the human subject, the irreducibility of meaning and value to data, and the need to retain interpretation, causal reasoning, and ethical deliberation in any use of data. HumAIsm insists that persons are more than the sum of their digital traces, that vulnerability and dignity must constrain optimisation, and that certain decisions---especially those affecting fundamental rights---require human judgment and contestability. This aligns with \citeauthor{Veliz2020}'s (\citeyear{Veliz2020}) view that privacy is collective and political: if data are concentrated in the hands of a few, democratic equality is undermined.

DatAIsm tends to treat platforms as neutral sensors of ``real behaviour,'' ignoring selection, mediation, and power. HumAIsm stresses that platforms are never neutral; their architectures and business models shape what is seen, measured, and optimized---and thus what becomes socially real. The choice of metrics, objectives and training data are normative decisions, not merely technical ones.

Data personalism gives this humanistic stance a normative anchor:

\begin{enumerate}
\item Personal data---especially those linked to intimacy, honour, image, and personality---are rights-laden and cannot be legitimately turned into pure commodities. This view resonates with the idea that personal data are extensions of personality, dignity and autonomy, not just alienable resources.

\item The apparent social acceptance of trading these rights (e.g., ``free'' services in exchange for intrusive data harvesting) reflects structural asymmetries and cognitive biases, not free and informed choice. DatAIsm normalises these arrangements by portraying them as efficient market exchanges; data personalism questions their legitimacy since there is only a very narrow and partial understanding of the consent given.

\item Digital traces cannot be interpreted as transparent manifestations of ``social reality'' but as partial, purposive, and contested signals that require interpretation and contextualization.
\end{enumerate}

The Digital Social Contract proposed here adopts this datapersonalist standpoint: personal data are not merely ``about'' persons; they emanate from them and remain bound up with their dignity. Any legitimate governance must respect this status. A dignity-centric digital social contract must therefore:

\begin{itemize}
\item Reject DatAIsm's reduction of persons to datapoints and resist technofeudal enclosures of digital life;
\item Affirm HumAIsm's commitment to persons as knowing, valuing and acting subjects;
\item Institutionalise data personalism through concrete rights, duties and design constraints that make respect for dignity non-optional in digital infrastructures.
\end{itemize}

In this sense, data sovereignty is not only a matter of jurisdiction or market power; it is a question of whose vision of the human person structures our digital institutions.

A further point follows naturally from this discussion. Personal data should not be treated, necessarily, as a factual representation of a person. A single datum, generated at a particular moment and under particular circumstances, might only capture an action, a situation or a context---not the full substance of an individual. Much of what is commonly inferred from data risks overlooking this contingency: data have, many times, a limited validity, they age, and they may reflect exceptions rather than stable traits. For this reason, any framework grounded in dignity must also safeguard the possibility for individuals to respond to, correct or contextualise the inferences or classifications made about them. A genuine right of reply is essential if we are to avoid mistaking momentary traces for enduring identity.

If DatAIsm treats persons as optimizable behavioral inputs, then a dignity-centric response must translate HumAIsm and data personalism into enforceable constraints. Next section states those constraints as a human-rights baseline for personal data.

\subsubsection{Personal Data as a Human Right: From Data Personalism to Human-Rights Protection}\label{sec:humanright}

This section advances towards the construction of a robust framework for limiting and protecting the collection and use of personal data, grounding them in a human-rights baseline and treating them as dignity-relevant manifestations of the person.

In contemporary digital infrastructures, personal data are socially operative representations that mediate identity and shape life chances. When datafied systems infer intimate attributes and govern access to opportunities, the normative question shifts from ``who owns data?'' to ``which forms of data power are compatible with human dignity and democratic life?'' \citep{EDPS2015, UNGA2014}.

The claim defended here is not that ``data'' in general constitutes a human right, but that personal data---understood data-personalistically as emanations of the person---should be protected under a fundamental-rights baseline, as the object of the right to the protection of personal data. This framing helps explain why consent alone or as a main basis cannot bear the entire ethical load and why some forms of extraction and use must be restricted even when individuals appear to ``agree'' under conditions of opacity, lock-in, platform dependence, or manipulative choice architecture \citep{Veliz2020, Zuboff2019}.

This diagnosis aligns with Solove's critique of ``privacy self-management'': modern privacy regimes have been dominated by an individual-control model that burdens persons with managing innumerable, opaque, and technically complex information relationships. The rise of AI makes the limits of this model especially visible, because individuals cannot realistically understand, assess, or continuously manage the downstream consequences of large-scale inference and automated decision systems. A dignity-centric digital social contract therefore requires a structural shift: privacy and data protection must rely less on formalistic control gestures and more on enforceable constraints over collection and use, meaningful organizational duties to prevent risks and harms, and accountability mechanisms that make remedies effective in practice \citep{Solove2025}.

The same structural critique applies to the recurring temptation to treat privacy notices as enforceable consumer contracts. \citet{Hartzog2026} argue that privacy notices persist in a ``weird twilight'' between mere policy description and binding agreement, and that contract doctrine has rarely empowered consumers in practice. More fundamentally, they contend that consumer contract law is an unsuitable foundation for privacy governance: it is insufficiently responsive to power disparities, overly focused on atomistic individual choice rather than collective and societal stakes, and built around conceptions of consent that become fictitious under conditions of scale, dependency, and opacity. In this framework, strengthening ``privacy-as-contract'' would not correct the imbalance; it would likely entrench it. Although that vision is grounded in the USA approach to privacy and it would not fully apply to the EU standards under the GDPR, it is true that a dignity-centric digital social contract therefore treats core protections as non-waivable and not reducible to private ordering through boilerplate terms.

Data personalism treats personal data as emanations of the person, but its normative force is not uniform across all informational artefacts. It applies with minimal friction to identifying/intimate and sensitive data, and to behavioural traces insofar as they enable inference of protected traits. Beyond individual cognitive limits, consent-centered contracting fails as a welfare mechanism in opaque, inference-driven ecosystems. When downstream uses are complex, multi-party, and difficult to foresee, individuals cannot reliably forecast consequences, and firms can profitably design take-it-or-leave-it choice architectures that convert dependence into ``agreement.'' In these conditions, observed acceptance does not reveal stable willingness-to-pay or willingness-to-accept for dignity-relevant protections; it reflects lock-in, bounded rationality, and bargaining asymmetry. Moreover, many privacy harms arise from social externalities: one person's disclosure changes inferences about others, and bilateral consent cannot internalize those effects. A rights-first baseline therefore functions as a non-waivable floor that prevents coercive ``privacy-for-access'' equilibria and shifts competition toward service quality and privacy-preserving innovation within Dignity-by-Design (hereafter DbD) constraints \citep{Acquisti2016, deHingh2018, Bergemann2022}.

A brief typology helps avoid a totalising claim: (i) identifying and intimate data; (ii) behavioural and contextual traces; (iii) inferred data and profiles (scores, predictions, model outputs), where power asymmetry and harm concentrate and where non-waivable limits are most salient; and (iv) relational/co-produced data (social graphs, household patterns), where governance cannot be reduced to individual consent because effects are collective. By contrast, the emanation claim is attenuated for data that are robustly anonymised or strongly aggregated such that persons are not reasonably identifiable (and thus fall outside the GDPR's scope); yet many ``aggregated'' or pseudonymised datasets remain linkable and inference-capable in practice, so the relevant test is realistic capacity to single out, link, infer, and decide across the data--information--knowledge chain.

Privacy and \textit{data protection} are related but distinct normative layers. \textit{Privacy} primarily protects the sphere of reserve and contextual integrity: it limits intrusions into what persons can keep undisclosed and under which expectations information may be shared. \textit{Data protection}, by contrast, is a governance regime for processing across the entire value chain, and therefore maps better onto the data--information--knowledge ladder: the most consequential harms often arise not at collection, but when raw traces are transformed into \textit{information} (combination, enrichment, communication) and especially into \textit{knowledge} (inferences, profiles, decision rules) that is applied to persons. The EU baseline makes this chain logic explicit: the core principles of Article 5 GDPR (lawfulness, purpose limitation, data minimisation, accuracy, integrity/confidentiality, accountability) discipline the \textit{data $\to$ information} stage, while the \textit{knowledge} stage triggers reinforced constraints around profiling and consequential decisions, notably Article 22 GDPR (limits on decisions based solely on automated processing with legal or similarly significant effects, with strict conditions and safeguards where exceptions apply).

The wider EU perimeter complements GDPR by constraining manipulative interfaces, gatekeeper data combination, and high-risk inferential deployments in the knowledge-as-power layer:

\begin{itemize}
\item DSA: prohibition of manipulative interface design (dark patterns) in online platforms.
\item DMA: constraints on gatekeepers' combination and cross-use of personal data across services absent GDPR-valid consent.
\item AI Act: bans of certain unacceptable-risk practices and governance duties for high-risk systems.
\end{itemize}

A human-rights claim for personal data can be grounded in at least four complementary academic lines:

\textit{Dignity and personality:} Personal data are intertwined with identity and personality; they are the informational ``surface'' through which honour, image, relationships, and self-presentation are formed and judged. A dignity-centric tradition therefore treats certain informational domains as inalienable or only conditionally waivable, requiring strong safeguards against commodification and abusive profiling \citep{UNESCO2000, EDPS2015}.

\textit{Enabling condition for other rights:} In a datafied society, effective privacy, freedom of expression and association, equal treatment, and due process increasingly depend on how personal data are collected, inferred, shared, and used. Human-rights protection is justified because personal data governance functions as an infrastructural precondition for the exercise of multiple rights in the digital age \citep{UNGA2014, Wong2023}.

\textit{Structural power asymmetry and market failure:} Personal data transactions occur under radical information asymmetries (invisible data flows, opaque inferences, complex ecosystems of intermediaries) and under conditions of platform dependence. This produces coercive ``privacy-for-access'' and ``privacy-for-cash'' dynamics and supports the need for non-negotiable floors of protection that do not collapse dignity into prices \citep{Zuboff2019, Varoufakis2023, Veliz2020}.

\textit{Collective and civic dimension:} Personal data practices generate collective externalities: profiling and recommender systems can amplify manipulation, polarization, and unequal life chances, eroding civic trust and democratic deliberation. A human-rights framing therefore requires duties not only toward individuals but also toward the public sphere, including transparency, accountability, and impact-based governance \citep{Susser2019, WorldBank2021, Wong2023}.

Taken together, these lines support a dignity-centric reading of data sovereignty: personal data governance must be rights-first, while markets and innovation remain permissible only within a non-waivable baseline. This provides the theoretical bridge to the principles and ``non-negotiable limits'' (see Annex~1) developed later in the paper and to DbD as a design obligation rather than an optional compliance layer.

This rights-first baseline does not deny that data generate economic value; it specifies the institutional conditions under which value creation is legitimate, distributively defensible, and compatible with dignity.

\subsubsection{Personal Data as a New Production Factor}\label{sec:productionfactor}

Data have become a new factor of production, alongside labor and capital, at the core of today's attention-driven economy; the explosive growth of data traffic and the training of ever larger AI models will further expand these volumes, making data an even more critical factor of production and significantly increasing its economic value. Treating personal data as a production factor describes their economic function; it does not settle their moral and legal status.

A growing body of academic and policy work explicitly frames data, and, particularly, personal data, as a distinct input in the production process of the digital economy. This is not only a metaphor (``data as the new oil''), but an analytic claim: data can be modeled as a factor that raises productivity and enables new products, especially when combined with computing infrastructure and skilled labor. In this line, \citet{Jones2020} draw a sharp conceptual distinction between ``ideas'' (blueprints/production functions) and ``data'' (an input used to produce, improve, and operationalize those ideas), arguing that ``data is a factor of production'' in modern growth and innovation dynamics.

The IMF adopts a closely aligned perspective. In its integrated overview of the economics of data, it identifies ``data as a factor of production'' as one of the two central economic functions of data: an input into the production of goods and services and a driver of innovation and efficiency through analysis and learning. Importantly, the IMF emphasizes that value extraction is not automatic: turning data into productive knowledge requires investment in storage, management, security, and---above all---complementary analytical capabilities and skilled labor \citep{CarriereSwallow2019}. This is particularly salient for personal data because the most economically valuable uses often involve profiling, prediction, and personalization.

Personal data exhibits properties that make it economically ``factor-like'' but also structurally different from labor, capital, or traditional intermediate inputs. First, personal data is frequently generated as a byproduct of consumption and social interaction (e.g., location traces, browsing, mobility, health signals), which creates a feedback loop: more users and more activity produce more data, which can improve services, attracting more users. Second, data is often nonrival (the same dataset can be used by multiple parties simultaneously), so the welfare gains from broader reuse can be large---but these gains collide with privacy costs, competitive hoarding incentives, and governance constraints \citep{Jones2020}. Third, the relevant economic question often shifts from ownership to access and control: because data can be copied and reused, the structure of permissions, licensing, and technical enforceability becomes central to how value is created and distributed \citep{Varian2018}.

This framing also helps clarify why the data economy tends toward concentration and asymmetry. Firms that accumulate large stocks of data can become more productive and can invest more aggressively, reinforcing their advantage (``data feedback loops''). Formal models of firm dynamics show how data accumulation can increase skewness in firm size and strengthen the position of data-rich incumbents, even when smaller ``data-savvy'' entrants can sometimes leapfrog if they manage to finance an initial phase of data accumulation and learning \citep{Farboodi2019}. In parallel, measurement work in macroeconomics increasingly treats data as an intangible asset (a storable input into production) that is only partially captured in standard statistics---highlighting the need for better accounting of data investments and their contribution to growth \citep{Corrado2022}.

However, recognizing personal data as a production factor should not imply reducing it to a mere commodity. In this paper's framework, personal data remains an emanation of the person---rights-laden and dignity-relevant---so the ``production factor'' lens must be paired with normative constraints and institutional design. Precisely because personal data is productive, distributive questions become unavoidable: Who captures the surplus created by personal data? Under what conditions is contribution genuinely voluntary and informed? What safeguards are required when productivity gains depend on intrusive inference? Proposals such as ``data as labor'' push in this direction by reframing user data generation as a contribution that may warrant bargaining power and compensation, especially given the monopsonistic structure of many digital markets \citep{ArrietaIbarra2018}.

\subsection{Proposed response: A Dignity-Centric Digital Social Contract}\label{sec:proposedresponse}

The preceding analysis clarifies the diagnosis of contemporary datafied societies. In response, we propose a Dignity-Centric Digital Social Contract as a framework for governing the data--information--knowledge chain under enforceable dignity constraints.

The normative foundation for a dignity-centric digital social contract draws on a longstanding tradition of social thought that predates the digital era. \citet{Francis2015}, in \textit{Laudato Si'}, identified the ``technocratic paradigm'' as a mode of governance in which efficiency and profit override the recognition of human dignity and the common good. In \textit{Fratelli Tutti}, Francis extended this critique to the digital sphere, arguing that hyperconnectivity does not inherently foster fraternity but may instead produce novel forms of exclusion and concentrate power among those who control technological infrastructures \citep{Francis2020}. These concerns were institutionalised in the Rome Call for AI Ethics, signed by the Holy See, Microsoft, IBM, the FAO, and the Italian Government, which established that artificial intelligence systems must satisfy requirements of transparency, inclusion, accountability, impartiality, reliability, and security in service of the human person \citep{RomeCall2020}. The convergence between this tradition of social thought and contemporary ethical reflection on technology provides the normative starting point from which the present proposal develops.

From a philosophical standpoint, the imperative to ground data governance in human dignity finds further support in the Kantian categorical imperative, which demands that persons never be treated merely as means \citep{Kant1997}. Applied to the data economy, this principle entails that personal data, as emanations of the person and extensions of personality, cannot be reduced to neutral inputs or alienable commodities \citep{deHingh2018, Floridi2016}. The European Data Protection Supervisor formalised this intuition by calling for ``a new digital ethics'' in which data and dignity are treated as inseparable \citep{EDPS2015}. More recently, \citet{Wong2023} has argued that data rights should be understood as a subset of human rights, reinforcing the claim that any governance framework for personal data must be evaluated against the standard of human dignity rather than mere economic efficiency.

The dignity-centric digital social contract responds to this analytical impasse by grounding data governance in the principle that personal data are emanations of the person, carriers of rights, and linked to human dignity, not neutral inputs or alienable commodities \citep{deHingh2018, Floridi2016}. This position, termed ``data personalism,'' operationalises the Kantian imperative within the specific architecture of the data economy. Its institutional translation proceeds through three mechanisms. First, the Dignity-by-Design principle requires that any system processing personal data incorporate ex ante safeguards for dignity, proportionality, and meaningful human oversight, establishing categorical prohibitions on the commercialisation of sensitive attributes and on automated decisions without human supervision. Second, redistributive instruments, including data dividends, data cooperatives, and fiduciary models, ensure that the value generated from personal data flows back to the communities and individuals who produce it, addressing the extraction dynamics identified by \citet{Srnicek2017}, \citet{Couldry2019}, and \citet{Varoufakis2023}. Third, a Pigouvian informational accounting framework internalises the externalities of mass data extraction by assigning measurable costs to informational harms, analogously to environmental taxation of pollutants \citep{GarridoMerchan2025, GarridoMerchan2026}.

Accordingly, our digital social contract approach integrates these economic insights while holding a rights-first baseline: productivity and innovation can be pursued, but only within architectures of meaningful control, proportionality, transparency, and fair benefit-sharing compatible with human dignity.

\section{Dignity-Centric Digital Social Contract. Core Dimensions}\label{sec:dimensions}

\begin{figure}[htbp]
\centering
\includegraphics[width=0.9\textwidth]{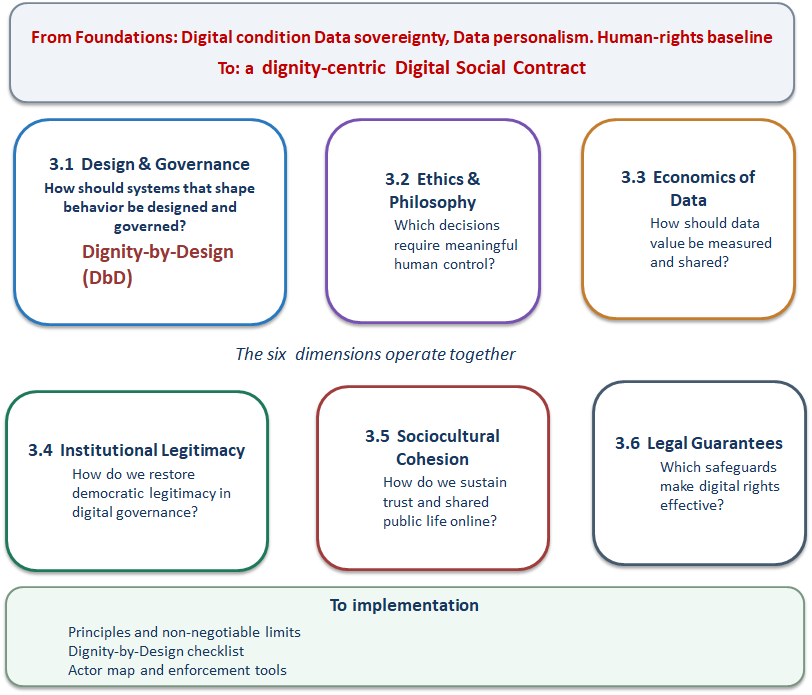}
\caption{Core dimensions of a dignity-centric digital contract.}
\label{fig:dimensions}
\end{figure}

This section develops the normative foundations of a dignity-centric digital social contract by articulating six complementary core dimensions that together define the scope of what such an order must protect and sustain (Figure~\ref{fig:dimensions}). Rather than treating personal data as a neutral commodity or mere economic input, the framework begins from the premise that data practices increasingly shape human agency, social inclusion, and life opportunities. The governance of digital infrastructures must therefore be anchored in the protection of human dignity as a foundational principle.

Throughout Section~\ref{sec:dimensions}, we use a consistent governance vocabulary: proportionality (necessity and minimisation), transparency (intelligibility to users), auditability (verifiability by qualified third parties), and remedy (effective contestation and redress). These terms link technical design choices to enforceable rights and legitimate institutional oversight.

\subsection{Technological Design and Governance: Dignity-by-Design}\label{sec:dbd}

The first dimension addresses the principles that should govern the design and deployment of systems that model and shape human behaviour. We propose extending the established \textit{privacy-by-design} approach into what we term Dignity-by-Design (DbD): the systematic embedding of transparency, auditability, explainability, security, fairness, and respect for personhood into all digital architectures and algorithms that process personal data.

A dignity-centric technological governance requires, in the first place, the deployment of Privacy-Enhancing Technologies (PETs)---including strong anonymisation, secure computation, and federated learning---to minimise personal data exposure wherever possible without compromising functionality. Equally important is the enforcement of legitimate purpose and data minimisation: personal data should be collected and used only for clear, justified purposes, avoiding gratuitous tracking or excessive data fusion beyond what is proportionate.

Users must be provided with meaningful control over their data through granular, easy-to-give and easy-to-revoke consent mechanisms that are free from manipulative \textit{dark patterns} (design features that nudge users into choices against their interests). Empirical evidence suggests that consent and opt-out mandates can induce strategic adaptation---shifts in tracking technologies, bidding strategies, and the composition of who remains trackable---so DbD must extend beyond collection to enforce traceability of data flows and constraints on downstream profiling and reuse \citep{Aridor2021}.

High-risk automated systems should be subject to mandatory Algorithmic Impact Assessments (AIAs) and, where applicable, Data Protection Impact Assessments (DPIAs), accompanied by independent audits. Critical algorithms should be transparent and traceable, with documented data sources and decision logic to enable ex-post review and remedies for harms.

Finally, DbD should be institutionalised through technical standards (such as interoperability and data portability by default) and certification mechanisms (ethical AI or data safety seals). Oversight bodies should be empowered to test algorithms for bias or risks and to enforce compliance, while public procurement can drive adoption by requiring DbD criteria in technology contracts. Data portability is often justified as pro-competition because it reduces switching costs, but welfare effects are not mechanically positive; they depend on how firms respond through pricing and bundling. The implication is not to abandon portability, but to implement it as part of a design-and-competition package: interoperability standards that make exported data usable, constraints on strategic bundling that degrades pass-through to users, and non-retaliation enforcement so portability is exercisable without service degradation \citep{Jeon2024, FlorezRamos2020}.

\subsection{Ethics and Philosophy: Human Oversight and Limits to Automation}\label{sec:ethics}

The second dimension concerns the decisions that must remain under meaningful human control. A dignity-centric approach requires preserving substantive human oversight---a genuine human-in-the-loop---for all consequential decisions affecting individuals' rights and well-being. Certain high-stakes domains, such as credit approval, healthcare, employment, criminal justice, or access to essential services, should not be left entirely to automated systems. Humans should retain final say or review, especially where decisions affect life opportunities or liberty.

This dimension also requires guaranteeing individual rights to explanation and contestation for algorithmic decisions. People should be able to understand why an automated decision was made about them and have accessible channels to challenge and seek redress for errors or unfair outcomes.

A further consideration is the acknowledgment that data and algorithms are not neutral. Digital traces carry \textit{methodological biases} and value judgments at each stage of the pipeline---collection, modelling, and interpretation. A dignity-centric approach recognises that no algorithmic decision-making is value-free, reinforcing the need for oversight and ethical constraints on AI systems.

Finally, key ethical safeguards should be embedded in digital systems as binding design requirements. These include nonmaleficence (do no harm)---for example, prohibiting the inference of sensitive personal attributes or manipulative micro-targeting without user consent; respect for autonomy, rejecting any coercive ``privacy-for-service'' trade-offs that force individuals to surrender dignity or privacy for access to basic services; and justice, including intersectional fairness---assessing algorithms for disparate impacts on vulnerable or marginalised groups (by gender, race, socioeconomic status, among others) and actively mitigating any biases or discrimination discovered.

\subsection{Economics of Data: Value, Redistribution, and Incentives}\label{sec:economics}

The third dimension addresses how value in a data-driven economy should be measured and fairly distributed. A central concern is the recognition of personal data as a key economic input and the correction of the current imbalance where individuals supply data while platforms capture the profits. Today's data economy functions in many respects like a monopsony: users' behavioural data are captured en masse, technology companies extract the insights and monetise them, benefits are concentrated, and privacy harms or externalities are spread across society. A dignity-centric digital social contract calls for mechanisms to rebalance this power and value distribution.

Any such rebalancing must begin with contextual data valuation. There is no single definition of ``data's value''---its worth depends on context (whose data, who uses it, for what purpose). Rather than reducing personal data to a simple commodity price, plural metrics are needed to measure data value at different scales. This means using both micro-level approaches (such as \textit{Data Shapley} value methods that attribute a model's performance gains to individual data contributions) and macro-level perspectives (treating data as an intangible asset in national accounts), as well as market-based and use-based valuations. A plurality of valuation methods prevents oversimplifying human worth into price tags.

However, valuation methods are not neutral descriptors; they operate as governance devices that can reshape bargaining power and market structure. The assetization literature emphasises that assets are made through practices that define boundaries, standardise metrics, and construct monetisation strategies that convert uncertain future gains into capitalised claims. In data economies, what is often valued is not an individual datum but the platform's capability to continuously produce prediction, targeting, and attention capture at scale---stabilised by measurement standards and engagement metrics that are legible to investors. Consequently, micro-valuation tools and market-price proxies should be treated as partial, context-specific signals useful for accountability and redistribution, not as foundations for turning dignity-relevant protections into alienable property rights. Willingness-to-pay and willingness-to-accept estimates can also be unstable when secondary trading is salient: evidence indicates that individuals experience disutility specifically from recipients' ability to profit via secondary-market monetisation, which can reduce participation and raise demanded prices. Personal data are, moreover, often ``social'' because inferences about one person inform beliefs about others; privacy design therefore implicates welfare conditions and externalities rather than only individual preference satisfaction \citep{Muniesa2017, Birch2020, Birch2021, Bergemann2022}.

Among the redistribution models proposed in the literature, the ``data as labour'' framework deserves particular attention. This approach treats data generation as a form of labour and explores mechanisms to compensate individuals for their data contributions, including direct payment for certain data, benefit provision in return, and support for collective bargaining through data cooperatives or data unions. Cooperative data platforms and unions enable individuals to pool their data and negotiate better terms (for example, revenue sharing or improved privacy safeguards), ensuring that those who generate data share in its benefits. Any such model must be designed carefully to avoid coercion or the exacerbation of inequalities---for instance, ensuring that those in economic need are not forced to ``sell'' their privacy at lower prices.

Several closely related proposals seek to correct data-market failures by strengthening property-like claims and explicit compensation for individual contributors, including ``data as labour'' arguments and broader calls for redesigned ``radical markets.'' These approaches share our diagnosis of bargaining asymmetry and missing prices, but they risk treating dignity-relevant protections as alienable in ways that can amplify regressivity when vulnerable groups face stronger short-term liquidity pressures. In our framework, monetary redistribution is therefore best treated as a constrained complement to a rights-first baseline (for instance, collective benefit-sharing funds or taxes on extractive practices), not as individualised sale of privacy protections. The key institutional question is not only how to pay, but which kinds of measurement and monetisation are permissible when the underlying resource remains rights-laden and socially externality-generating \citep{ArrietaIbarra2018, Posner2018, deHingh2018, Bergemann2024}.

Two additional mechanisms warrant consideration. First, redistribution towards the public interest: some have proposed \textit{data dividends} or public data trusts that return a portion of data monetisation profits to the community. While experimental, these ideas align with the principle that the wealth generated from collective data should benefit society at large, not just a few corporations. Second, the internalisation of externalities: introducing Pigouvian taxes or fees for data practices that carry privacy risks or cause social harms (for instance, a levy on invasive targeted advertising or on algorithms that amplify harmful content), using the revenue to fund privacy protections, digital literacy, or other public goods. By pricing in negative externalities such as loss of privacy or discrimination, firms are pushed to innovate in more privacy-preserving and fair directions.

Finally, this economic dimension requires investment in digital literacy and empowerment across society, so that individuals understand how their data are used and what their value is. Many people undervalue their personal data or underestimate privacy risks due to limited awareness (a phenomenon akin to a Dunning--Kruger effect in data literacy). Educational initiatives can empower users to make informed choices, demand better terms, or participate in data stewardship initiatives. An informed public is better equipped to insist on fair compensation and accountability in the data economy.

\subsection{Political-Institutional Legitimacy: Multi-Actor, Multi-Level Governance}\label{sec:political}

The fourth dimension addresses the conditions for legitimate, democratic governance of digital platforms and data at both national and global levels. A first requirement is the reaffirmation of the role of the state and public institutions in governing digital infrastructure. This includes establishing specialised digital regulatory agencies equipped with audit powers and the authority to sanction abuses by powerful technology platforms. Governments should regain the capacity to set rules for data and AI systems in the public interest, without stifling innovation but also without ceding oversight to private actors.

Equally important is the embedding of democratic deliberation into digital governance through inclusive policymaking. This can take the form of digital citizens' assemblies or participatory forums on AI and data policy, where diverse stakeholders---citizens, experts, civil society---can debate and guide major decisions. Such mechanisms ensure that the ``social contract'' online is shaped through open dialogue and has legitimacy, rather than being unilaterally set by corporations.

Civil society and collective action also require institutional support. This means bolstering digital rights NGOs, consumer advocacy groups, academic watchdogs, and emerging data unions that represent individuals' data interests. Such groups can hold companies accountable, advocate for users' rights, and provide an organised counterweight to corporate power in data governance.

Since data flows and digital platforms cross borders, legitimacy further demands international cooperation. Progress towards international agreements on minimum standards for digital rights and data protection, and coordinated enforcement against transnational technology abuses, is essential. Multi-level governance---from local to global---is needed so that no single country or company can erode fundamental rights without consequences.

Finally, existing legal frameworks can serve as building blocks for legitimacy. Instruments like the EU's GDPR, as well as newer laws such as the Data Governance Act and Data Act, offer partial regulatory blueprints. They introduce ideas such as neutral data intermediaries, safe data-sharing spaces, user data portability rights, and restrictions on unfair contract terms in data exchange. These can be expanded and adopted beyond the EU to form a baseline for trustworthy digital governance across jurisdictions.

\subsection{Sociocultural Cohesion: Rebuilding the Digital Commons}\label{sec:sociocultural}

The fifth dimension concerns the conditions under which data practices and digital platforms can sustain, rather than erode, social trust, inclusion, and the conditions for shared public life. A first priority is the mitigation of echo chambers and the fragmentation of public discourse by algorithmic curation. This requires greater transparency in how social media and content platforms personalise information feeds. Key algorithms---including recommender systems---should be auditable by independent researchers to assess their societal impact. By illuminating how content is ranked or targeted, it becomes possible to address the creation of filter bubbles and to reduce the amplification of sensationalism and polarisation.

Content moderation regimes should be grounded in human rights principles, upholding freedom of expression while protecting users from harm. Platforms should adopt moderation policies with clear rules against hate speech, harassment, and misinformation, together with safeguards against over-removal or censorship of legitimate speech. Oversight boards or external auditors can help ensure moderation is fair, accountable, and culturally sensitive, thus maintaining trust in online spaces.

Investment in digital literacy and inclusion is equally essential. Widespread media and data literacy programmes are needed so that users can navigate the online world critically and safely. This includes educating users---especially young people, who may be ``digital natives'' yet overconfident about their skills---on how to identify misinformation, protect their privacy, and understand algorithmic biases. Strengthening individual competencies helps build collective resilience against manipulation and fosters more informed, cohesive communities online.

Rebuilding shared civic spaces on the internet is also necessary. This could involve funding for quality journalism, open-source social networks, libraries of knowledge, and community-driven content platforms. A healthy digital commons means people encounter diverse perspectives and trustworthy information, which underpins social cohesion in the digital age. In this sense, recent work frames the (digital) commons explicitly as a techno-political response to platform oligarchies, connecting commons governance to democratic objectives rather than consumer surplus alone \citep{Perperidis2026}.

Finally, the principles of data justice should be applied to ensure that data practices promote equity and represent all groups. This involves focusing on redistribution (fair sharing of data's benefits), recognition (respect for different communities' digital identities and norms), and representation (inclusive decision-making in data governance). Policies should be evaluated for their impact on social inclusion: do they reduce or increase digital divides? A dignity-centric contract strives to make digital society more just and inclusive, countering the corrosive effects of surveillance capitalism and emerging ``technofeudal'' dynamics that treat people as mere data sources.

\subsection{Legal-Regulatory Guarantees: Effective Digital Rights}\label{sec:legal}

The sixth dimension identifies the legal and institutional safeguards required to ensure that digital infrastructures remain compatible with fundamental rights and human dignity. A foundational step is the formal recognition of key digital rights as fundamental rights. These include rights to privacy, personal identity, data protection, non-discrimination in algorithmic decisions, the right to explanation for AI outcomes, data portability, and the \textit{right to disconnect} from pervasive surveillance. Enshrining such rights in constitutions or international human rights frameworks would cement the principle that personal data are an extension of the person, deserving of the highest legal protection.

The framework further requires imposing fiduciary-like duties of loyalty and care on data controllers and AI providers. Just as doctors or lawyers must act in their clients' best interests, major platforms should have a duty of loyalty to their users, meaning they must not use personal data in ways that betray users' trust or well-being. This could be achieved through legislation that defines data fiduciaries obligated to handle personal data with loyalty, care, and confidentiality. This shift is also supported by recent arguments that privacy governance should not be built on consumer contract doctrine: where bargaining power is structurally absent, purported ``agreements'' become unilateral rule-setting rather than meaningful consent. In such contexts, consumer-protection safeguards and fiduciary-like obligations are more appropriate vehicles for constraining data power than private ordering through boilerplate terms \citep{Hartzog2026}.

Individuals must have effective means to enforce their digital rights. This includes collective redress mechanisms (allowing groups of users to band together to challenge systemic abuses, akin to class-action lawsuits for data breaches or privacy violations) and independent digital ombudspersons or regulators who can mediate complaints. Regulators should have the ability to levy substantial fines or even criminal penalties on organisations that egregiously violate data rights or engage in practices harmful to human dignity.

Clear legal red lines must also be set on activities that are incompatible with human dignity. These include, for instance, banning the unchecked trade in sensitive personal data (such as health, biometric, or intimate information) for profiling or marketing; prohibiting fully automated decision-making in critical areas of life without human oversight or consent; and outlawing manipulative design practices that exploit cognitive biases to extract data. Such categorical prohibitions guard against the worst abuses of data-driven technology.

The operationalisation of these guarantees can build on existing and emerging regulatory frameworks. The EU's General Data Protection Regulation (GDPR) already anchors data protection in fundamental rights. Building on this, laws like the Data Governance Act and Data Act introduce trusted data intermediaries, facilitate data sharing for the public good (data altruism), and mandate fairness in data access (for example, preventing IoT device manufacturers from hoarding user data). Adopting and strengthening similar laws across jurisdictions will create a more robust legal architecture, ensuring that technological development does not outrun the protection of human dignity and rights. One market-design proposal operationalises this institutional role via licensed personal-data intermediaries acting as users' exclusive agents and bound by fiduciary duties, underscoring that even ``market'' approaches require regulator-set standards and enforceable constraints rather than consent alone \citep{Bergemann2024}.

A final and increasingly pressing concern is the treatment of publicly available data and scraping. Large-scale scraping and aggregation of ``public'' personal data should be treated as regulated collection, not as a categorical exemption from privacy protection. In the age of AI, many privacy interests attach to data that is technically public but practically obscure, context-bounded, dispersed, and not ordinarily recorded at scale. A dignity-centric contract should therefore safeguard practical obscurity and prohibit indiscriminate collection of personal data from the internet for arbitrary purposes, subjecting such practices to purpose limitation, proportionality, and enforceable accountability \citep{Solove2025}.

Due to length reasons, we leave for Annex~1 the operational blueprint through which the six core dimensions can be implemented in terms of 5 pillars (institutional levers), 7 principles (general governance criteria) and 6 non-negotiable limits (categorical prohibitions).

\section{Discussion and Future Research Agenda}\label{sec:discussion}

This paper argues that digital life now requires an explicit dignity-centric Digital Social Contract---one that treats personal data as dignity-relevant manifestations of the person and that governs the full data--information--knowledge value chain. The central motivation is structural: the current technological disruption has given rise to contemporary digital infrastructures that convert ordinary traces into profiles, predictions, and consequential decisions at scale, under conditions of opacity, cross-border fragmentation, and platform dependence. In this environment, governance failures do not arise merely from isolated privacy breaches, but from the cumulative effects of inference-driven systems that redistribute power, shape life chances, and erode the conditions for autonomy, equality, and democratic legitimacy.

In this sense, we claim that we are not only facing a technological disruption, but a \textit{human disruption.} Since \citeauthor{Christensen1997}'s (\citeyear{Christensen1997}) seminal work, technological disruption has generally been interpreted through the lens of markets and firms \citep[cf.][]{Millar2018}. However, the capacity of current technological disruption to create entirely new contexts or radically transform existing ones extends far beyond markets. This disruption is altering the status quo of social, cultural, and political orders, rendering what were once considered stable pillars increasingly fluid. In this sense, current digital technological disruption is not merely economic but anthropological. It is a \textit{human disruption}, a disruption that is provoking a change on the status quo of the human being in society, including changes in the size and depth of the public sphere, and changes in the capacity of individuals to appear and act within it as moral and relational subjects \citep{Arendt1973}.

Against this situation, two opposing responses may be given. A first, datAIsm-inspired stance denies the need for political reflection: algorithmic systems, continuously fed by data, are assumed to operate as an ``automated invisible hand'' capable of governing society efficiently, pixelating human life into a supposedly more realistic representation of individuals and collectives. On this view, institutional design becomes unnecessary, as data-driven optimization would replace the historically human foundations of freedom, self-realization, and participatory democracy with governance grounded in ``reality,'' understood as data \citep{Sadin2020}. We reject this position since justice, equality, dignity and legitimacy are not technical variables that can be derived from datasets, and society cannot be treated as an engineering problem. A second response insists that disruption requires deliberate institutional choice: societies must actively decide which political and social structures should endure, which will be displaced by algorithmic systems, and on what normative grounds. In this human-centered view that we defend, technology is a tool rather than an autonomous agent; it lacks logos, truth-claims, and the authority to govern persons. Reducing human action, meaning, and relationships to predictable ``behavior'' is therefore conceptually and morally flawed. Technological innovation must remain subordinated to human agency, and disruption should be governed through explicit reflection on its implications for dignity, democracy, and the common good.

This is why our dignity-centric approach is not an optional ethical add-on but a necessary institutional response to the logic of datafied societies.

In this sense, the subsequent emergence of OpenClaw and Moltbook offers an early, practice-based confirmation of the paper's central claim: once the principles of data sovereignty are translated into real workflows and technical artefacts, the need for \textit{verifiable} dignity constraints, traceability, proportionality, meaningful control, and enforceable red lines, stops being theoretical and becomes operationally unavoidable. In that sense, OpenClaw and Moltbook function as ex post ``proof in practice'' that a dignity-centric approach must be built into systems by design (and not appended as compliance language), reinforcing the case for institutionalised auditability and meaningful oversight across the data--information--knowledge chain \citep{EDPS2015, Mantelero2022}.

This section reframes the main implications as discussion points and as a publishable research agenda. We define the research and implementation frontier. It aligns the agenda with the six governance dimensions developed in Section~\ref{sec:dimensions} and with the escalation logic implied by the DIKW ladder: as systems move from traces to profiles and to consequential decisions, duties must intensify and become auditable.

\subsection{Contributions, limitations, and contested frontiers}\label{sec:whatclarifies}

Our contribution is twofold. First, we defend a data-personalist foundation: personal data are not neutral inputs or ordinary commodities but rights-laden emanations of the person, and therefore require a rights-first baseline that cannot be reduced to consent, contract, or market exchange as the main basis for the processing of personal data. Second, we translate that baseline into an operational program through Dignity-by-Design (DbD), extending privacy-by-design into a broader set of design and governance obligations---purpose proportionality, continuous control, transparency and explainability, auditability, traceability, security, fairness, and respect for personhood. The framework is structured through six core dimensions---technological design and governance; human oversight and limits to automation; value, redistribution, and incentives; legitimacy and multi-actor governance; sociocultural cohesion and the digital commons; and legal-regulatory guarantees---because the data crisis is simultaneously technological, ethical, economic, political-institutional, sociocultural, and legal. A key implication is that partial fixes predictably propagate failure elsewhere: technical safeguards without enforceability remain cosmetic; legal rights without usable control remain formal; and economic redistribution without non-waivable limits risks pricing dignity.

\begin{figure}[htbp]
\centering
\includegraphics[width=0.85\textwidth]{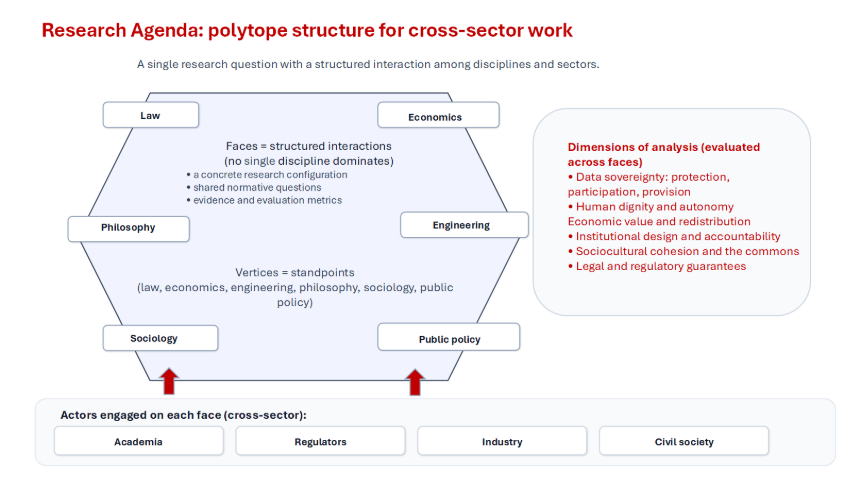}
\caption{Polytope Structure for cross-sector work.}
\label{fig:polytope}
\end{figure}

This multidimensional architecture also clarifies why our argument is best understood through a polytope logic rather than a linear policy sequence (Figure~\ref{fig:polytope}). The ``polytope'' is not a metaphor for complexity alone, but an epistemic and institutional design principle: each dimension represents a vertex, while the faces represent the interaction effects where governance succeeds or fails. For example, privacy-enhancing technologies cannot deliver dignity if market incentives reward extraction; valuation and redistribution schemes cannot be legitimate if they weaken rights; and regulatory guarantees cannot be effective if auditability is technically impossible. The polytope structure therefore implies that implementation requires coordinated advances across disciplines and sectors, rather than isolated progress within a single field. In this sense, a dignity-centric Digital Social Contract raises substantive questions where reasonable cross-discipline debate may persist, and this is best framed as research and implementation frontiers rather than as weaknesses of the approach.

The main contested question may be the regulatory framework defended in this paper, towards which a well-articulated tradition of liberal and libertarian thought may raise principled objections. \citet{Hayek1945}, in ``The Use of Knowledge in Society,'' argued that decentralised market mechanisms process dispersed information far more efficiently than any centralised planning authority, a thesis that contemporary data-market advocates extend to personal data flows: regulatory intervention, on this view, distorts the price signals through which data finds its most productive allocation. \citet{Hayek1960} further argued in \textit{The Constitution of Liberty} that individual freedom requires limiting government intervention to the enforcement of general rules, not the redistribution of specific outcomes, a principle that, applied to data governance, would preclude ex ante prohibitions on data uses or mandatory redistribution schemes such as data dividends.

\citet{Friedman1962}, in \textit{Capitalism and Freedom}, provided the canonical argument that economic freedom is both an end in itself and a necessary condition for political freedom, and that government regulation of voluntary exchanges, including data transactions freely consented to, constitutes a paternalistic infringement on individual autonomy. \citet{Nozick1974}, in \textit{Anarchy, State, and Utopia}, extended this logic to its most rigorous conclusion: any pattern of distribution imposed by the state violates the rights of individuals who have acquired holdings through legitimate voluntary transfers, so that even well-intentioned redistribution of data-derived value constitutes an unjust taking. \citet{Rothbard1973}, in \textit{For a New Liberty}, and \citet{Mises1949}, in \textit{Human Action}, pushed the libertarian argument to its anarcho-capitalist limit, contending that all functions of the state, including the enforcement of property rights, can and should be supplied by voluntary market arrangements.

In the contemporary technology sector, these positions find expression in the ``effective accelerationism'' (e/acc) movement, whose proponents argue that innovation should be maximally accelerated without regulatory interference \citep{Andreessen2023}. \citet{Thiel2009} articulated the underlying tension explicitly: ``I no longer believe that freedom and democracy are compatible,'' suggesting that democratic governance necessarily constrains the dynamism required for technological progress. \citet{Land2012}, in \textit{The Dark Enlightenment}, provided the philosophical architecture for this position, proposing an unregulated hypercapitalism that celebrates the elimination of democratic mediation and the progressive displacement of human agency by autonomous technological processes, an explicit anti-humanism in which concentration of power is not merely tolerated but welcomed as an index of systemic fitness.

A third accelerationist current---anarchist accelerationism---proposes that technology should dismantle all hierarchical structures, including the state, through decentralised cryptographic protocols \citep{May1992, Hughes1993}. Its institutional expressions include DAOs and blockchain governance \citep{DeFilippi2018}. However, eliminating institutional mediation does not eliminate power asymmetries: it displaces them onto code architectures where protocol founders exercise de facto sovereignty without accountability \citep{Lessig1999, Schneider2022}. A dignity-centred contract values decentralisation as a design principle while insisting that decentralisation without enforceable dignity constraints produces a different topology of domination, not freedom.

The danger of this convergence between classical libertarianism and technological anti-humanism is empirically identifiable. The ``data feedback loops'' documented by \citet{Farboodi2019} and \citet{Jones2020} demonstrate that data accumulation reinforces the competitive advantages of dominant platforms exponentially, replicating and amplifying the tendency toward concentration and centralisation that \citet{Marx1990} identified as inherent to the functioning of capital. When personal data operate as a factor of production without their generators participating in the value created \citep{ArrietaIbarra2018}, the Landian logic of the Dark Enlightenment materialises in what \citet{Varoufakis2023} terms ``technofeudalism'' and \citet{Zuboff2019} terms ``surveillance capitalism'': a regime in which a small number of platforms extract rents over digital infrastructures while users are reduced to informational serfs. \citet{Couldry2019} have characterised this dynamic as ``data colonialism,'' arguing that global platforms replicate historical patterns of colonial extraction by appropriating the raw material of human life itself.

Further than that, some other questions may be contested, among which we highlight the following:

First, the operational scope of non-negotiable limits may launch an interesting debate. Annex~1 articulates six categorical red lines---commercialisation of especially protected attributes; fully automated consequential decisions without meaningful human intervention; intrusive collection without necessity and minimisation; dark patterns and lock-in; opaque transfers to third parties; and retaliation for exercising rights. These limits reflect the core diagnosis that consent fails structurally under dependency, manipulation, and opaque inference. Yet the practical challenge is to define operational tests for when a system crosses a dignity threshold, particularly in cases of sensitive inference from proxies, multi-model pipelines, and ``human oversight'' that is procedural rather than substantive. Future work should therefore develop measurable, regulator-ready criteria for dignity violations, including robust indicators of inferential sensitivity, contestability, and meaningful intervention.

Second, the paper's economic dimension opens a necessary but delicate research agenda on valuation and benefit-sharing. Recognizing personal data as a production factor clarifies why distributive questions are unavoidable, but it does not justify commodification. Data dividends, licensing schemes, and ``data as labor'' proposals can contribute to justice when they operate as complements to rights, yet they risk pricing dignity when they function as substitutes for non-waivable protections. Future research should therefore evaluate redistribution mechanisms through bounded real-world pilots, with explicit attention to regressivity, coercion, and heterogeneous impacts on vulnerable groups. In parallel, valuation models should incorporate informational externalities and intersectional risk, since many harms arise socially and relationally, as one person's disclosure alters inferences about others, and bilateral consent cannot internalize those effects.

Third, the feasibility of DbD in practice is an empirical and institutional question, not merely a normative one. DbD presupposes audit capacity, documentation discipline, and technical standards that enable verification. A key risk is performative compliance, as organisations may treat documentation as a symbolic layer rather than as evidence of binding constraints on inference and decision-making. This concern is sharpened by the techno-regulatory imaginary identified in privacy-engineering debates, where rights risk being collapsed into design requirements while deeper issues of power, incentives, and enforceability are displaced. Future work should therefore focus on institutionalizing DbD as an auditable governance method, standardizing traceability artefacts (data sheets, model cards, system cards), strengthening independent review capacity, and developing scalable impact and bias testing methods aligned with dignity constraints.

Fourth, the institutional dimension highlights the need to study governance models beyond individual rights and state-centric regulation. Data cooperatives, trusted intermediaries, and emerging ``data spaces'' promise new forms of participation and bargaining power, but their success conditions remain poorly understood. Comparative research should examine real-world intermediaries and cooperative structures, identify failure modes (capture, low adoption, weak incentives), and evaluate which regulatory models best sustain dignity-centric outcomes under different market structures. A closely related avenue concerns the design of user-facing data agents and consent interfaces: innovation is needed not only in privacy-enhancing computation but also in human-centered governance tools that make rights usable and non-retaliatory in practice.

Fifth, the sociocultural dimension remains under-measured despite being central to the legitimacy of the contract. Data-driven platforms reshape everyday life by normalizing extraction and commodification of intimate behaviours, potentially corroding ordinary trust and the conditions for self-development. These dynamics appear particularly acute for adolescents, whose developmental vulnerability is amplified by algorithmically optimized environments, and whose harms are unevenly distributed along intersectional lines. Future work should therefore measure polarization, trust, and civic engagement under different algorithmic curation regimes, and should develop robust approaches to digital and data literacy---including systematic biases in perceived privacy risk and data value among ``digital natives.'' In this sense, the Digital Social Contract is not only a framework for individual rights but also a governance project for the digital commons.

Finally, global legitimacy beyond the EU baseline remains unresolved. While GDPR and the broader EU perimeter provide a partial blueprint for a rights-first settlement, data power is transnational and enforcement capacity diverges sharply across jurisdictions. Future research should map convergences and divergences in digital rights and data sovereignty, propose minimum global principles for cross-border data governance, and explore credible enforcement mechanisms that avoid externalizing harms to weaker regulatory regimes. Relatedly, the collective dimension of data sovereignty remains institutionally underdeveloped: current legal and technical frameworks still focus mainly on individual remedies, while many harms are population-level and relational. Advancing collective remedies, shared governance for common datasets, and public-interest data institutions is therefore a priority for the next stage of dignity-centric governance.

Taken together, these discussion points show that the Digital Social Contract proposed here is not a closed doctrine but a structured program: it specifies a rights-first baseline, operationalizes it through DbD, and identifies the interdisciplinary work required to make dignity constraints enforceable in real systems. The polytope structure clarifies why progress must be coordinated across design, law, economics, institutions, and sociocultural dynamics. A digital society that accepts permanent opacity, unilateral rule-setting, and scalable behavioural control cannot remain compatible with democratic legitimacy or with the equal dignity of persons. The core challenge is therefore not whether innovation should continue, but how innovation can be governed under a renewed settlement that makes dignity, rights, and accountability non-optional across the data--information--knowledge chain. In this sense, dignity-centric innovation is a field that warrants sustained scholarly attention going forward.

\subsection{A structured research program across themes}\label{sec:researchprogram}

These frontiers reinforce the central claim of the paper: a dignity-centric Digital Social Contract is not a closed normative statement but an integrated research and governance programme. Each theme links to the paper's core dimensions (technology, ethics, economics, legitimacy, sociocultural cohesion, and law) and to the DIKW escalation logic.

A first discussion frontier concerns the operationalisation of dignity thresholds and non-waivable limits. The paper defends categorical constraints on certain data practices, even where nominal consent is present, because consent systematically fails under conditions of opacity, dependency, and inference-based expansion \citep{Solove2024, Solove2025}. Yet the practical challenge is to translate these limits into regulator-ready criteria that can be applied across sectors and technical architectures. Future research must therefore specify how dignity thresholds can be detected and evidenced in practice. This includes developing methods to identify proxy-based sensitive inference, where systems infer protected attributes without explicit collection, and to distinguish between meaningful human intervention and purely procedural oversight in hybrid decision systems \citep[e.g.,][]{Mantelero2022, deHingh2018, Barocas2016, Susser2019}. Advancing this line of work requires combining legal analysis, system auditing, and empirical study of real-world pipelines to define measurable indicators of dignity-relevant risk along the data--information--knowledge chain.

A second discussion frontier lies in the governance of DbD as an auditable method. DbD extends privacy-by-design by embedding dignity-relevant constraints into system architectures, documentation, and organisational processes. However, its effectiveness depends on whether it produces artefacts that can sustain enforcement rather than merely symbolic compliance. Without institutionalisation, DbD risks being absorbed into a techno-regulatory imaginary that treats rights as design features while leaving underlying power relations intact \citep{Rommetveit2022, Mantelero2022}.

A third discussion frontier concerns the shift of data governance toward general-purpose AI and the training-data pipeline. As foundation models restructure the digital ecosystem, the effective locus of data power increasingly lies upstream, in large-scale scraping, dataset curation, and model training \citep{Solove2025, Couldry2019, Mejias2024}. This development challenges governance approaches that focus primarily on downstream applications or individual consent events \citep{Pasquale2015, Bratton2016}. Future research must therefore examine how dignity constraints apply to the collection and reuse of ``public'' personal data at scale, how training-data documentation and provenance affect accountability in practice, and what technical and legal mechanisms make deletion meaningful once models have been trained. Comparative case studies of closed and open model ecosystems can clarify how risks and responsibilities are distributed across supply chains, and whether current governance tools adequately reach the knowledge-production layer where inference power concentrates.

A fourth frontier relates to data sovereignty beyond individual consent \citep{Abbas2024, Cohen2013, Solove2024, Scholz2016}. Our paper conceptualises data sovereignty as a multidimensional construct combining protection, participation, and provision, but this triad requires institutional embodiments that shift bargaining power without creating new concentration points. Future research should therefore evaluate the real-world performance of data cooperatives, trusted intermediaries, and emerging data spaces, identifying success conditions and failure modes such as capture, low adoption, or weak incentives. Closely related is the study of fiduciary-like intermediaries and personal data agents, which promise to mediate data governance on behalf of users. Assessing whether such institutions enhance agency or merely repackage dependence requires comparative fieldwork and governance analysis across jurisdictions and sectors.

A fifth discussion frontier concerns the interaction between dignity-centric rights and market structure \citep{Jones2020, Farboodi2019, Varian2018}. Recognising personal data as a production factor makes distributive questions unavoidable, but it does not settle how value should be shared without commodifying dignity \citep{ArrietaIbarra2018, Zuboff2019, Varoufakis2023}. Future research must therefore examine when rights-based constraints intensify market concentration and when they enable entry and innovation, particularly in platform markets characterised by data feedback loops. This includes studying how interoperability and portability affect bargaining power after their implementation, and which fiscal or fee-based instruments can internalise data externalities without legitimising extraction \citep{Jeon2024, Frey2024}. Empirical work that combines structural modelling with evaluation of policy shocks is needed to ensure that dignity-centric governance is paired with competition remedies rather than inadvertently reinforcing incumbency.

A sixth frontier addresses sociocultural cohesion and child protection as governance outcomes rather than externalities \citep{Tufekci2014, Susser2019, vanDijck2018}. The legitimacy of a digital social contract depends not only on individual rights but also on public-sphere effects such as trust, polarisation, and inclusion. Yet these outcomes remain difficult to measure under current access regimes. Future research should therefore develop robust methodologies to assess how recommender systems and interface designs affect civic engagement and social fragmentation, and which interventions reduce harms to minors without expanding surveillance \citep{Veliz2020}.

A seventh discussion frontier concerns global legitimacy and cross-border enforcement \citep{Pohle2020, Floridi2024, Chander2023}. While the framework aligns closely with the European rights architecture, data power is transnational and regulatory capacity remains uneven. Future research must therefore identify a minimal dignity-protective floor that is politically feasible across jurisdictions, and evaluate enforcement tools that do not rely solely on extraterritorial reach. Comparative analysis of global instruments and mutual-assistance mechanisms can clarify how a dignity-centric contract might operate beyond regional boundaries without collapsing into lowest-common-denominator standards.

The proposal to develop a Dignity of Data Index raises a further discussion frontier \citep{Ghorbani2019, Bendechache2023}. Indices can guide behaviour and inform policy, but they also risk being gamed or mis-specified \citep{Muniesa2017, Birch2020}. Future research must therefore determine which indicators predict real dignity outcomes rather than procedural compliance, how protection, participation, and provision should be weighted across sectors, and how index governance can preserve independence and contestability.

A related but distinct research line concerns cybersecurity as an integral component of dignity-centric governance. Data breaches, system vulnerabilities, and weak security architectures frequently translate into dignity harms by exposing intimate information, enabling coercion or discrimination, and eroding trust in digital infrastructures. Yet cybersecurity is often analysed separately from data protection and AI governance. Future research should therefore investigate how secure-by-design principles can be systematically integrated into Dignity-by-Design frameworks, treating resilience, access control, and incident response as dignity-relevant obligations rather than purely technical safeguards.

Framing these challenges explicitly as research lines clarifies how dignity-centric innovation can move from principle to practice without sacrificing democratic legitimacy or human dignity.

\subsection{Dignity-centric Innovation as a distinct research line}\label{sec:dignityinnovation}

The proposed Digital Social Contract entails that dignity can guide innovation toward architectures that are auditable, contestable, and socially legitimate, rather than functioning only as an ex post limit that slows technological development. Under this view, innovation is not assessed only by gains in efficiency or performance, but by whether new systems expand human capabilities while remaining compatible with a rights-first baseline across the full data--information--knowledge chain. This reframing motivates dignity-centric innovation as a distinct and necessary research line.

Dignity-centric innovation requires a redefinition of what counts as progress in data-driven systems. Rather than equating innovation with accuracy, scale, or efficiency, it should be understood as dignity-centric when it expands functional capability while reducing dignitarian risk as systems move from data collection to inference and decision-making. This framing shifts attention away from privacy at the point of collection toward the stages where informational power concentrates: profiling, prediction, ranking, and automated or semi-automated decisions \citep{Barocas2016, Mantelero2022, Solove2025}. Conceptually, this implies prioritising technical approaches that deliver services without enabling high-risk inference, such as privacy-preserving personalisation, federated and secure computation, on-device processing, and identity mechanisms based on minimal, purpose-bound data \citep{Acquisti2016, Goldfarb2012, Goldfarb2019}. Equally important are organisational and policy conditions that make such designs viable in practice, including procurement requirements, product governance structures, audit readiness, incident-response capacity, and clear accountability allocation along data and model supply chains \citep{EDPS2015, OECD2024}.

At the methodological level, dignity-centric innovation demands evaluation tools that move beyond standard performance metrics. Accuracy and efficiency provide limited insight into the social and normative consequences of inference-driven systems. A central challenge is to measure inferential expansion: the gap between what individuals knowingly disclose and what systems infer through aggregation, enrichment, and modelling \citep{MayerSchonberger2013, Susser2019}. Evaluation frameworks must also test whether human oversight substantively affects outcomes, rather than functioning as a procedural add-on \citep{Mantelero2022}. In parallel, researchers and regulators must assess group-level and public-sphere harms, such as discrimination or manipulation, without escalating surveillance through additional sensitive data collection. This motivates the development of dignity-centric assessment protocols that complement DPIAs and AIAs, as well as benchmark suites for inference risk, contestability, and non-retaliation suitable for independent audit.

At the institutional level, dignity-centric innovation requires environments that enable experimentation while keeping rights binding. Regulatory uncertainty can otherwise produce either excessive risk aversion or informal norm-breaking. Future research should therefore examine regulatory sandbox models that support learning without diluting non-negotiable limits, and the role of public procurement in rewarding auditable dignity-centric designs rather than formal compliance alone \citep{OECD2024}. Shared infrastructures---standards, certification, audit tooling, and secure research-access regimes---are also critical to lowering entry barriers for smaller firms and civic innovators.

Together, these elements justify treating dignity-centric innovation as a distinct research programme that aligns technological development with human dignity, democratic legitimacy, and effective accountability under conditions of data-driven power.

\section{Conclusion}\label{sec:conclusion}

Digital infrastructures mediate core domains of social life while they record behaviour, infer traits, and steer choices through ranking, recommendation, and automated decision systems. These infrastructures operate across borders and through complex supply chains. As a result, the classic social contract framework faces a crisis of territoriality, enforcement, public norm-setting, and legitimacy. Rights may exist in law, but people lack practical leverage, intelligibility, and effective remedies.

This paper presents a dignity-centric Digital Social Contract grounded in data personalism. It treats personal data as rights-laden emanations of the person and as a basis for a human-rights claim to protection. From this starting point, the paper reframes data sovereignty as a multidimensional settlement that integrates protection, participation, and provision. It also clarifies why governance must extend beyond the collection stage to the information and knowledge stages, where profiling, inference, and consequential decisions concentrate power.

The operational contribution is a coherent programme that connects normative commitments to implementable duties. Six governance dimensions organise the agenda across technology, ethics, economics, political-institutional legitimacy, sociocultural cohesion, and law. Dignity-by-Design translates these dimensions into system requirements that organisations can specify, test, and audit. Non-negotiable limits set categorical red lines where consent and contractual boilerplate cannot legitimate extraction, manipulation, or high-stakes automation. Redistributive mechanisms and informational externality accounting complement this rights-first baseline without turning dignity into a price. The framework also invites coordinated action by regulators, firms, researchers, and civil society.

A viable digital future requires societies that can verify claims, enforce constraints, and provide remedy at scale. Future work should develop tests for proxy-based sensitive inference, standards for meaningful human intervention, and methods that measure public-sphere harms without expanding surveillance. Progress on these fronts would allow innovation to proceed within dignity constraints, restoring legitimacy to digital governance and strengthening the conditions for autonomy, equality, and civic trust.

\bibliographystyle{apalike}
\bibliography{references}

\begin{thebibliography}{}

\bibitem[Abbas et~al., 2024]{Abbas2024}
Abbas, A.~E., van Velzen, T., Ofe, H., van~de Kaa, G., Zuiderwijk, A., and
  de~Reuver, M. (2024).
\newblock Beyond control over data: Conceptualizing data sovereignty from a
  social contract perspective.
\newblock {\em Electronic Markets}, 34(20).

\bibitem[Ackoff, 1989]{Ackoff1989}
Ackoff, R.~L. (1989).
\newblock From data to wisdom.
\newblock {\em Journal of Applied Systems Analysis}, 16:3--9.

\bibitem[Acquisti et~al., 2016]{Acquisti2016}
Acquisti, A., Taylor, C., and Wagman, L. (2016).
\newblock The economics of privacy.
\newblock {\em Journal of Economic Literature}, 54(2):442--492.

\bibitem[Al-Rodhan, 2014]{AlRodhan2014}
Al-Rodhan, N. R.~F. (2014).
\newblock The social contract 2.0: Big data and the need to guarantee privacy
  and civil liberties.
\newblock {\em Harvard International Review}.

\bibitem[Andreessen, 2023]{Andreessen2023}
Andreessen, M. (2023).
\newblock The techno-optimist manifesto.
\newblock Andreessen Horowitz.

\bibitem[Arendt, 1973]{Arendt1973}
Arendt, H. (1973).
\newblock {\em The Origins of Totalitarianism}.
\newblock Houghton Mifflin Harcourt.

\bibitem[Aridor et~al., 2021]{Aridor2021}
Aridor, G., Che, Y.~K., and Salz, T. (2021).
\newblock The effect of privacy regulation on the data industry: Empirical
  evidence from {GDPR}.
\newblock In {\em Proceedings of the 22nd ACM Conference on Economics and
  Computation}, pages 93--94.

\bibitem[Arrieta-Ibarra et~al., 2018]{ArrietaIbarra2018}
Arrieta-Ibarra, I., Goff, L., Jim\'{e}nez-Hern\'{a}ndez, D., Lanier, J., and
  Weyl, E.~G. (2018).
\newblock Should we treat data as labor? {Moving} beyond ``free''.
\newblock {\em AEA Papers and Proceedings}, 108:38--42.

\bibitem[B{\"a}ckstrand et~al., 2017]{Backstrand2017}
B{\"a}ckstrand, K., Kuyper, J.~W., Linn\'{e}r, B.~O., and L{\"o}vbrand, E.
  (2017).
\newblock Non-state actors in global climate governance: from {Copenhagen} to
  {Paris} and beyond.
\newblock {\em Environmental Politics}, 26(4):561--579.

\bibitem[Barocas and Selbst, 2016]{Barocas2016}
Barocas, S. and Selbst, A.~D. (2016).
\newblock Big data's disparate impact.
\newblock {\em California Law Review}, 104(3):671--732.

\bibitem[Bastani, 2019]{Bastani2019}
Bastani, A. (2019).
\newblock {\em Fully Automated Luxury Communism: A Manifesto}.
\newblock Verso Books.

\bibitem[Bendechache et~al., 2023]{Bendechache2023}
Bendechache, M., Attard, J., Ebiele, M., and Brennan, R. (2023).
\newblock A systematic survey of data value: Models, metrics, applications and
  research challenges.
\newblock {\em IEEE Access}, 11:104966--104983.

\bibitem[Bergemann and Bonatti, 2024]{Bergemann2024}
Bergemann, D. and Bonatti, A. (2024).
\newblock Data, competition, and digital platforms.
\newblock {\em American Economic Review}, 114(8):2553--2595.

\bibitem[Bergemann et~al., 2022]{Bergemann2022}
Bergemann, D., Bonatti, A., and Gan, T. (2022).
\newblock The economics of social data.
\newblock {\em The RAND Journal of Economics}, 53(2):263--296.

\bibitem[Bergemann et~al., 2018]{Bergemann2018}
Bergemann, D., Bonatti, A., and Smolin, A. (2018).
\newblock The design and price of information.
\newblock {\em American Economic Review}, 108(1):1--48.

\bibitem[Berkelaar, 2014]{Berkelaar2014}
Berkelaar, B.~L. (2014).
\newblock Cybervetting, online information, and personnel selection: New
  transparency expectations and the emergence of a digital social contract.
\newblock {\em Management Communication Quarterly}, 28(4):479--506.

\bibitem[Birch et~al., 2021]{Birch2021}
Birch, K., Cochrane, D.~T., and Ward, C. (2021).
\newblock Data as asset? {The} measurement, governance, and valuation of
  digital personal data by {Big Tech}.
\newblock {\em Big Data \& Society}, 8(1).

\bibitem[Birch and Muniesa, 2020]{Birch2020}
Birch, K. and Muniesa, F. (2020).
\newblock {\em Assetization: Turning Things into Assets in Technoscientific
  Capitalism}.
\newblock MIT Press.

\bibitem[Boucher and Kelly, 1994]{Boucher1994}
Boucher, D. and Kelly, P., editors (1994).
\newblock {\em The Social Contract from {Hobbes} to {Rawls}}.
\newblock Routledge, London.

\bibitem[Boyd and Crawford, 2012]{Boyd2012}
Boyd, d. and Crawford, K. (2012).
\newblock Critical questions for big data.
\newblock {\em Information, Communication \& Society}, 15(5):662--679.

\bibitem[Bratton, 2016]{Bratton2016}
Bratton, B.~H. (2016).
\newblock {\em The Stack: On Software and Sovereignty}.
\newblock MIT Press.

\bibitem[Brynjolfsson and McAfee, 2014]{Brynjolfsson2014}
Brynjolfsson, E. and McAfee, A. (2014).
\newblock {\em The Second Machine Age: Work, Progress, and Prosperity in a Time
  of Brilliant Technologies}.
\newblock W. W. Norton.

\bibitem[Cammatte, 1973]{Cammatte1973}
Cammatte, J. (1973).
\newblock {\em The Wandering of Humanity}.
\newblock Black \& Red.
\newblock Trans.\ F.\ Perlman.

\bibitem[Cardelli et~al., 2020]{Cardelli2020}
Cardelli, L., Orgad, L., Shahaf, G., Shapiro, E., and Talmon, N. (2020).
\newblock Digital social contracts: A foundation for an egalitarian and just
  digital society.
\newblock In {\em CEUR Workshop Proceedings}, volume 2781, pages 51--60.

\bibitem[Carri\`{e}re-Swallow and Haksar, 2019]{CarriereSwallow2019}
Carri\`{e}re-Swallow, Y. and Haksar, V. (2019).
\newblock The economics and implications of data: An integrated perspective.
\newblock IMF Departmental Paper 19/16, International Monetary Fund.

\bibitem[Cassese, 2003]{Cassese2003}
Cassese, S. (2003).
\newblock The age of administrative reforms.
\newblock In {\em Governing Europe}, pages 128--138.

\bibitem[Castells, 1996]{Castells1996}
Castells, M. (1996).
\newblock {\em The Rise of the Network Society}.
\newblock Blackwell.

\bibitem[Castro~Fern\'{a}ndez, 2025]{CastroFernandez2025}
Castro~Fern\'{a}ndez, R. (2025).
\newblock What is the value of data? {A} theory and systematization.
\newblock {\em ACM/IMS Journal of Data Science}.

\bibitem[Chander and Sun, 2023]{Chander2023}
Chander, A. and Sun, H., editors (2023).
\newblock {\em Data Sovereignty: From the Digital Silk Road to the Return of
  the State}.
\newblock Oxford University Press.

\bibitem[Chandler, 2015]{Chandler2015}
Chandler, D. (2015).
\newblock A world without causation: Big data and the coming of age of
  posthumanism.
\newblock {\em Millennium}, 43(3):833--851.

\bibitem[Choucri, 2012]{Choucri2012}
Choucri, N. (2012).
\newblock {\em Cyberpolitics in International Relations}.
\newblock MIT Press.

\bibitem[Christensen, 1997]{Christensen1997}
Christensen, C.~M. (1997).
\newblock {\em The Innovator's Dilemma: When New Technologies Cause Great Firms
  to Fail}.
\newblock Harvard Business School Press, Boston, MA.

\bibitem[Clinton and Perera, 2016]{Clinton2016}
Clinton, L. and Perera, D., editors (2016).
\newblock {\em The Cybersecurity Social Contract: Implementing a Market-Based
  Model for Cybersecurity}.
\newblock Internet Security Alliance.

\bibitem[Cohen, 2013]{Cohen2013}
Cohen, J.~E. (2013).
\newblock What privacy is for.
\newblock {\em Harvard Law Review}, 126(7):1904--1933.

\bibitem[Corrado et~al., 2022]{Corrado2022}
Corrado, C., Haskel, J., Iommi, M., and Jona-Lasinio, C. (2022).
\newblock Measuring data as an asset: Framework, methods and preliminary
  estimates.
\newblock OECD Economics Department Working Papers 1731, OECD Publishing.

\bibitem[Couldry and Mejias, 2019]{Couldry2019}
Couldry, N. and Mejias, U.~A. (2019).
\newblock Data colonialism: Rethinking big data's relation to the contemporary
  subject.
\newblock {\em Television \& New Media}, 20(4):336--349.

\bibitem[Couture and Toupin, 2019]{Couture2019}
Couture, S. and Toupin, S. (2019).
\newblock What does the notion of ``sovereignty'' mean when referring to the
  digital?
\newblock {\em New Media \& Society}, 21(10):2305--2322.

\bibitem[De~Filippi and Wright, 2018]{DeFilippi2018}
De~Filippi, P. and Wright, A. (2018).
\newblock {\em Blockchain and the Law: The Rule of Code}.
\newblock Harvard University Press.

\bibitem[de~Hingh, 2018]{deHingh2018}
de~Hingh, A. (2018).
\newblock Some reflections on dignity as an alternative legal concept in data
  protection regulation.
\newblock {\em German Law Journal}, 19(5):1269--1290.

\bibitem[Deleuze and Guattari, 1972]{Deleuze1972}
Deleuze, G. and Guattari, F. (1972).
\newblock {\em L'Anti-Oedipe: Capitalisme et schizophr\'{e}nie}.
\newblock Les \'{E}ditions de Minuit.

\bibitem[Dyer-Witheford, 2015]{DyerWitheford2015}
Dyer-Witheford, N. (2015).
\newblock {\em Cyber-Proletariat: Global Labour in the Digital Vortex}.
\newblock Pluto Press.

\bibitem[{European Data Protection Supervisor}, 2015]{EDPS2015}
{European Data Protection Supervisor} (2015).
\newblock Opinion 4/2015: Towards a new digital ethics---data, dignity and
  technology.
\newblock Technical report.

\bibitem[{European Parliamentary Research Service}, 2021]{EPRS2021}
{European Parliamentary Research Service} (2021).
\newblock Artificial intelligence at {EU} borders: Overview of applications and
  key issues.
\newblock Technical report, European Parliament.

\bibitem[Farboodi et~al., 2019]{Farboodi2019}
Farboodi, M., Mihet, R., Philippon, T., and Veldkamp, L. (2019).
\newblock Big data and firm dynamics.
\newblock {\em AEA Papers and Proceedings}, 109:38--42.

\bibitem[Ferguson, 2017]{Ferguson2017}
Ferguson, N. (2017).
\newblock {\em The Square and the Tower: Networks and Power, from the
  Freemasons to Facebook}.
\newblock Penguin Press.

\bibitem[Fl\'{o}rez-Ramos and Blind, 2020]{FlorezRamos2020}
Fl\'{o}rez-Ramos, E. and Blind, K. (2020).
\newblock Data portability effects on data-driven innovation of online
  platforms: Analyzing {Spotify}.
\newblock {\em Telecommunications Policy}, 44(9):102026.

\bibitem[Floridi, 2016]{Floridi2016}
Floridi, L. (2016).
\newblock Faultless responsibility: On the nature and allocation of moral
  responsibility for distributed moral actions.
\newblock {\em Philosophical Transactions of the Royal Society A},
  374(2083):20160112.

\bibitem[Floridi et~al., 2024]{Floridi2024}
Floridi, L., Fratini, S., Hine, E., Novelli, C., and Roberts, H. (2024).
\newblock Digital sovereignty: A descriptive analysis and a critical evaluation
  of existing models.
\newblock {\em Digital Society}, 3(59).

\bibitem[Francis, 2015]{Francis2015}
Francis (2015).
\newblock {\em Laudato Si': On Care for Our Common Home}.
\newblock Libreria Editrice Vaticana.

\bibitem[Francis, 2020]{Francis2020}
Francis (2020).
\newblock {\em Fratelli Tutti: On Fraternity and Social Friendship}.
\newblock Libreria Editrice Vaticana.

\bibitem[Fratini et~al., 2024]{Fratini2024}
Fratini, S., Hine, E., Novelli, C., Roberts, H., and Floridi, L. (2024).
\newblock Digital sovereignty: A descriptive analysis and a critical evaluation
  of existing models.
\newblock {\em Digital Society}, 3(3):59.

\bibitem[Frey and Presidente, 2024]{Frey2024}
Frey, C.~B. and Presidente, G. (2024).
\newblock Privacy regulation and firm performance: Estimating the {GDPR} effect
  globally.
\newblock {\em Economic Inquiry}, 62(3):1074--1089.

\bibitem[Friedman, 1962]{Friedman1962}
Friedman, M. (1962).
\newblock {\em Capitalism and Freedom}.
\newblock University of Chicago Press.

\bibitem[Garrido-Merch\'{a}n, 2025]{GarridoMerchan2025}
Garrido-Merch\'{a}n, E.~C. (2025).
\newblock An information-theoretic intersectional data valuation theory.
\newblock {\em arXiv preprint arXiv:2507.14742}.

\bibitem[Garrido-Merch\'{a}n, 2026]{GarridoMerchan2026}
Garrido-Merch\'{a}n, E.~C. (2026).
\newblock Intersectional data and the social cost of digital extraction: A
  {Pigouvian} surcharge.
\newblock {\em arXiv preprint arXiv:2601.08574}.

\bibitem[Ghorbani and Zou, 2019]{Ghorbani2019}
Ghorbani, A. and Zou, J. (2019).
\newblock Data {Shapley}: Equitable valuation of data for machine learning.
\newblock In {\em Proceedings of the 36th International Conference on Machine
  Learning}, pages 2242--2251. PMLR.

\bibitem[Ghosh and Scott, 2018]{Ghosh2018}
Ghosh, D. and Scott, B. (2018).
\newblock Digital deceit: The technologies behind precision propaganda on the
  internet.

\bibitem[Gillespie, 2010]{Gillespie2010}
Gillespie, T. (2010).
\newblock The politics of `platforms'.
\newblock {\em New Media \& Society}, 12(3):347--364.

\bibitem[Gitelman, 2013]{Gitelman2013}
Gitelman, L., editor (2013).
\newblock {\em Raw Data Is an Oxymoron}.
\newblock MIT Press.

\bibitem[Goldfarb and Tucker, 2012]{Goldfarb2012}
Goldfarb, A. and Tucker, C. (2012).
\newblock Shifts in privacy concerns.
\newblock {\em American Economic Review}, 102(3):349--353.

\bibitem[Goldfarb and Tucker, 2019]{Goldfarb2019}
Goldfarb, A. and Tucker, C. (2019).
\newblock Digital economics.
\newblock {\em Journal of Economic Literature}, 57(1):3--43.

\bibitem[Goldfarb and Tucker, 2011]{Goldfarb2011}
Goldfarb, A. and Tucker, C.~E. (2011).
\newblock Privacy regulation and online advertising.
\newblock {\em Management Science}, 57(1):57--71.

\bibitem[Hardt and Negri, 2000]{Hardt2000}
Hardt, M. and Negri, A. (2000).
\newblock {\em Empire}.
\newblock Harvard University Press.

\bibitem[Hardt and Negri, 2004]{Hardt2004}
Hardt, M. and Negri, A. (2004).
\newblock {\em Multitude: War and Democracy in the Age of Empire}.
\newblock Penguin Press.

\bibitem[Hartzog and Solove, 2026]{Hartzog2026}
Hartzog, W. and Solove, D.~J. (2026).
\newblock Privacy as contract?
\newblock {\em Available at SSRN}.

\bibitem[Hayek, 1945]{Hayek1945}
Hayek, F.~A. (1945).
\newblock The use of knowledge in society.
\newblock {\em American Economic Review}, 35(4):519--530.

\bibitem[Hayek, 1960]{Hayek1960}
Hayek, F.~A. (1960).
\newblock {\em The Constitution of Liberty}.
\newblock University of Chicago Press.

\bibitem[{High Level Expert Group on Artificial Intelligence},
  2019]{HLEGAI2019}
{High Level Expert Group on Artificial Intelligence} (2019).
\newblock Ethics guidelines for trustworthy {AI}.
\newblock Technical report, European Commission.

\bibitem[Hobbes, 2012]{Hobbes2012}
Hobbes, T. (2012).
\newblock {\em Leviathan}.
\newblock Cambridge University Press.
\newblock Original work published 1651.

\bibitem[Hughes, 1993]{Hughes1993}
Hughes, E. (1993).
\newblock A cypherpunk's manifesto.

\bibitem[Hummel et~al., 2021]{Hummel2021}
Hummel, P., Braun, M., Tretter, M., and Dabrock, P. (2021).
\newblock Data sovereignty: A review.
\newblock {\em Big Data \& Society}, 8(1):1--17.

\bibitem[Jeon and Menicucci, 2024]{Jeon2024}
Jeon, D.~S. and Menicucci, D. (2024).
\newblock Data portability and competition: Can data portability increase both
  consumer surplus and profits?
\newblock {\em European Journal of Law and Economics}, 57(1):145--162.

\bibitem[Jones and Tonetti, 2020]{Jones2020}
Jones, C.~I. and Tonetti, C. (2020).
\newblock Nonrivalry and the economics of data.
\newblock {\em American Economic Review}, 110(9):2819--2858.

\bibitem[Kant, 1997]{Kant1997}
Kant, I. (1997).
\newblock {\em Groundwork of the Metaphysics of Morals}.
\newblock Cambridge University Press.
\newblock Trans.\ M.\ Gregor. Original work published 1785.

\bibitem[Kitchin, 2014]{Kitchin2014}
Kitchin, R. (2014).
\newblock Big data, new epistemologies and paradigm shifts.
\newblock {\em Big Data \& Society}, 1(1).

\bibitem[Kukutai and Taylor, 2016]{Kukutai2016}
Kukutai, T. and Taylor, J., editors (2016).
\newblock {\em Indigenous Data Sovereignty: Toward an Agenda}.
\newblock ANU Press.

\bibitem[Land, 2012]{Land2012}
Land, N. (2012).
\newblock The dark enlightenment.
\newblock Online essay series.

\bibitem[Lessig, 1999]{Lessig1999}
Lessig, L. (1999).
\newblock {\em Code and Other Laws of Cyberspace}.
\newblock Basic Books.

\bibitem[Liaropoulos, 2020]{Liaropoulos2020}
Liaropoulos, A. (2020).
\newblock A social contract for cyberspace.
\newblock {\em Journal of Information Warfare}, 19(2):1--11.

\bibitem[Locke, 1988]{Locke1988}
Locke, J. (1988).
\newblock {\em Two Treatises of Government}.
\newblock Cambridge University Press.
\newblock Original work published 1689.

\bibitem[Mantelero, 2022]{Mantelero2022}
Mantelero, A. (2022).
\newblock {\em Beyond Data: Human Rights, Ethical and Social Impact Assessment
  in {AI}}.
\newblock Springer Nature.

\bibitem[Marcuse, 1964]{Marcuse1964}
Marcuse, H. (1964).
\newblock {\em One-Dimensional Man: Studies in the Ideology of Advanced
  Industrial Society}.
\newblock Beacon Press.

\bibitem[Marx, 1973]{Marx1973}
Marx, K. (1973).
\newblock {\em Grundrisse: Foundations of the Critique of Political Economy}.
\newblock Penguin Books.
\newblock Original work published 1858.

\bibitem[Marx, 1990]{Marx1990}
Marx, K. (1990).
\newblock {\em Capital: A Critique of Political Economy, Volume {I}}.
\newblock Penguin Classics.
\newblock Original work published 1867.

\bibitem[Mason, 2015]{Mason2015}
Mason, P. (2015).
\newblock {\em PostCapitalism: A Guide to Our Future}.
\newblock Allen Lane.

\bibitem[May, 1992]{May1992}
May, T.~C. (1992).
\newblock The crypto anarchist manifesto.

\bibitem[Mayer-Sch{\"o}nberger and Cukier, 2013]{MayerSchonberger2013}
Mayer-Sch{\"o}nberger, V. and Cukier, K. (2013).
\newblock {\em Big Data: A Revolution That Will Transform How We Live, Work,
  and Think}.
\newblock Eamon Dolan/Houghton Mifflin Harcourt.

\bibitem[McChesney, 2007]{McChesney2007}
McChesney, R.~W. (2007).
\newblock {\em Communication Revolution: Critical Junctures and the Future of
  Media}.
\newblock New Press.

\bibitem[Mejias and Couldry, 2024]{Mejias2024}
Mejias, U.~A. and Couldry, N. (2024).
\newblock Data grab: The new colonialism of big tech and how to fight back.
\newblock In {\em Data Grab}. University of Chicago Press.

\bibitem[Millar et~al., 2018]{Millar2018}
Millar, C., Lockett, M., and Ladd, T. (2018).
\newblock Disruption: technology, innovation and society.
\newblock {\em Technological Forecasting and Social Change}, 129:254--260.

\bibitem[Mises, 1949]{Mises1949}
Mises, L.~v. (1949).
\newblock {\em Human Action: A Treatise on Economics}.
\newblock Yale University Press.

\bibitem[Moulier-Boutang, 2011]{MoulierBoutang2011}
Moulier-Boutang, Y. (2011).
\newblock {\em Cognitive Capitalism}.
\newblock Polity Press.

\bibitem[Muniesa et~al., 2017]{Muniesa2017}
Muniesa, F., Doganova, L., Ortiz, H., Pina-Stranger, A., Paterson, F.,
  Bourgoin, A., Ehrenstein, V., Juven, P.-A., Pontille, D., Sarac-Lesavre, B.,
  and Yon, G. (2017).
\newblock {\em Capitalization: A Cultural Guide}.
\newblock Presses des Mines, Paris.

\bibitem[Nani, 2023]{Nani2023}
Nani, A. (2023).
\newblock Data as intangible assets.
\newblock {\em Meditari Accountancy Research}.

\bibitem[Negri, 1991]{Negri1991}
Negri, A. (1991).
\newblock {\em Marx Beyond Marx: Lessons on the Grundrisse}.
\newblock Autonomedia.

\bibitem[Nozick, 1974]{Nozick1974}
Nozick, R. (1974).
\newblock {\em Anarchy, State, and Utopia}.
\newblock Basic Books.

\bibitem[{OECD}, 2024]{OECD2024}
{OECD} (2024).
\newblock {AI} governance and digital trust.
\newblock Technical report, OECD Publishing.

\bibitem[O'Neil, 2016]{ONeil2016}
O'Neil, C. (2016).
\newblock {\em Weapons of Math Destruction: How Big Data Increases Inequality
  and Threatens Democracy}.
\newblock Crown.

\bibitem[O'Reilly, 2005]{OReilly2005}
O'Reilly, T. (2005).
\newblock What is {Web} 2.0: Design patterns and business models for the next
  generation of software.
\newblock {\em Communications \& Strategies}, 65(1):17--37.

\bibitem[{Parliament and Council of the European Union}, 2016]{GDPR2016}
{Parliament and Council of the European Union} (2016).
\newblock Regulation ({EU}) 2016/679 ({General Data Protection Regulation}).
\newblock Official Journal of the European Union, L 119.

\bibitem[Pasquale, 2015]{Pasquale2015}
Pasquale, F. (2015).
\newblock {\em The Black Box Society: The Secret Algorithms That Control Money
  and Information}.
\newblock Harvard University Press.

\bibitem[Perperidis and Tsekeris, 2026]{Perperidis2026}
Perperidis, G. and Tsekeris, C. (2026).
\newblock Imagining techno-political alternatives: the (digital) commons as a
  response to tech oligarchies.
\newblock {\em Science as Culture}, pages 1--15.

\bibitem[Pigou, 1920]{Pigou1920}
Pigou, A.~C. (1920).
\newblock {\em The Economics of Welfare}.
\newblock Macmillan.

\bibitem[Pohle and Thiel, 2020]{Pohle2020}
Pohle, J. and Thiel, T. (2020).
\newblock Digital sovereignty.
\newblock {\em Internet Policy Review}, 9(4):1--19.

\bibitem[{Pontifical Academy for Life} et~al., 2020]{RomeCall2020}
{Pontifical Academy for Life}, {Microsoft}, {IBM}, {FAO}, and {Italian
  Government} (2020).
\newblock Rome call for {AI} ethics.
\newblock The Vatican.

\bibitem[Posner and Weyl, 2018]{Posner2018}
Posner, E.~A. and Weyl, E.~G. (2018).
\newblock {\em Radical Markets: Uprooting Capitalism and Democracy for a Just
  Society}.
\newblock Princeton University Press.

\bibitem[Qiao et~al., 2022]{Qiao2022}
Qiao, T., Li, Y., Zhao, Y., Tan, C., and Zhang, P. (2022).
\newblock Exploration of theory and method of studying digital governance
  pattern.
\newblock {\em Bulletin of Chinese Academy of Sciences (Chinese Version)},
  37(10):1365--1375.

\bibitem[Rommetveit and Van~Dijk, 2022]{Rommetveit2022}
Rommetveit, K. and Van~Dijk, N. (2022).
\newblock Privacy engineering and the techno-regulatory imaginary.
\newblock {\em Social Studies of Science}, 52(6):853--877.

\bibitem[Rothbard, 1973]{Rothbard1973}
Rothbard, M.~N. (1973).
\newblock {\em For a New Liberty: The Libertarian Manifesto}.
\newblock Macmillan.

\bibitem[Rousseau, 1997]{Rousseau1997}
Rousseau, J.-J. (1997).
\newblock {\em The Social Contract and Other Later Political Writings}.
\newblock Cambridge University Press.
\newblock Original work published 1762.

\bibitem[Sadin, 2020]{Sadin2020}
Sadin, E. (2020).
\newblock {\em L'\`{E}re de l'individu tyran. La fin d'un monde commun}.
\newblock Grasset.

\bibitem[Scherer et~al., 2016]{Scherer2016}
Scherer, A.~G., Rasche, A., Palazzo, G., and Spicer, A. (2016).
\newblock Managing for political corporate social responsibility: New
  challenges and directions for {PCSR} 2.0.
\newblock {\em Journal of Management Studies}, 53(3):273--298.

\bibitem[Schneider, 2022]{Schneider2022}
Schneider, N. (2022).
\newblock {\em Governable Spaces: Democratic Design for Online Life}.
\newblock University of California Press.

\bibitem[Scholz, 2016]{Scholz2016}
Scholz, T. (2016).
\newblock {\em Platform Cooperativism: Challenging the Corporate Sharing
  Economy}.
\newblock Rosa Luxemburg Stiftung.

\bibitem[Shadmy, 2019]{Shadmy2019}
Shadmy, T. (2019).
\newblock The new social contract: {Facebook}'s community and our rights.
\newblock {\em Boston University International Law Journal}, 37:307.

\bibitem[Shannon, 1948]{Shannon1948}
Shannon, C.~E. (1948).
\newblock A mathematical theory of communication.
\newblock {\em Bell System Technical Journal}, 27(3--4):379--423, 623--656.

\bibitem[Shin and Shin, 2025]{Shin2025}
Shin, E.~Y. and Shin, D. (2025).
\newblock Trustworthy {AI} and the governance of misinformation: policy design
  and accountability in the fact-checking system.
\newblock {\em Transforming Government: People, Process and Policy}, pages
  1--19.

\bibitem[Solove, 2024]{Solove2024}
Solove, D.~J. (2024).
\newblock Murky consent: an approach to the fictions of consent in privacy law.
\newblock {\em Boston University Law Review}, 104:593.

\bibitem[Solove, 2025]{Solove2025}
Solove, D.~J. (2025).
\newblock {\em On Privacy and Technology}.
\newblock Oxford University Press.

\bibitem[Spraul and H{\"o}fert, 2021]{Spraul2021}
Spraul, K. and H{\"o}fert, A. (2021).
\newblock Governance for sustainability: Patterns of regulation and
  self-regulation in the {German} wine industry.
\newblock {\em Sustainability}, 13(6):3140.

\bibitem[Srinivasan and Ghosh, 2023]{Srinivasan2023}
Srinivasan, R. and Ghosh, D. (2023).
\newblock A new social contract for technology.
\newblock {\em Policy \& Internet}, 15(1):117--132.

\bibitem[Srnicek, 2017]{Srnicek2017}
Srnicek, N. (2017).
\newblock {\em Platform Capitalism}.
\newblock John Wiley \& Sons.

\bibitem[Srnicek and Williams, 2015]{Srnicek2015}
Srnicek, N. and Williams, A. (2015).
\newblock {\em Inventing the Future: Postcapitalism and a World Without Work}.
\newblock Verso Books.

\bibitem[Starr, 2004]{Starr2004}
Starr, P. (2004).
\newblock {\em The Creation of the Media: Political Origins of Modern
  Communications}.
\newblock Basic Books.

\bibitem[Susser et~al., 2019]{Susser2019}
Susser, D., Roessler, B., and Nissenbaum, H. (2019).
\newblock Online manipulation: Hidden influences in a digital world.
\newblock {\em Georgetown Law Technology Review}, 4(1):1--45.

\bibitem[Thiel, 2009]{Thiel2009}
Thiel, P. (2009).
\newblock The education of a libertarian.
\newblock Cato Unbound.

\bibitem[Tufekci, 2014]{Tufekci2014}
Tufekci, Z. (2014).
\newblock Engineering the public: Big data, surveillance and computational
  politics.
\newblock {\em First Monday}, 19(7).

\bibitem[{UNESCO}, 2000]{UNESCO2000}
{UNESCO} (2000).
\newblock Infoethics 2000: Protecting human dignity in the digital age.

\bibitem[{United Nations General Assembly}, 2014]{UNGA2014}
{United Nations General Assembly} (2014).
\newblock The right to privacy in the digital age.
\newblock A/RES/68/167.

\bibitem[van Dijck, 2014]{vanDijck2014}
van Dijck, J. (2014).
\newblock Datafication, dataism \& dataveillance: Big data between scientific
  paradigm and ideology.
\newblock {\em Surveillance \& Society}, 12(2):197--208.

\bibitem[van Dijck et~al., 2018]{vanDijck2018}
van Dijck, J., Poell, T., and de~Waal, M. (2018).
\newblock {\em The Platform Society: Public Values in an Online World}.
\newblock Oxford University Press.

\bibitem[Varian, 2018]{Varian2018}
Varian, H.~R. (2018).
\newblock Artificial intelligence, economics, and industrial organization.
\newblock NBER Working Paper 24839, National Bureau of Economic Research.

\bibitem[Varoufakis, 2023]{Varoufakis2023}
Varoufakis, Y. (2023).
\newblock {\em Technofeudalism: What Killed Capitalism}.
\newblock Bodley Head.

\bibitem[V\'{e}liz, 2020]{Veliz2020}
V\'{e}liz, C. (2020).
\newblock {\em Privacy Is Power: Why and How You Should Take Back Control of
  Your Data}.
\newblock Melville House.

\bibitem[Vitiea and Lim, 2019]{Vitiea2019}
Vitiea, K. and Lim, S. (2019).
\newblock Voluntary environmental collaborations and corporate social
  responsibility in {Siem Reap} city, {Cambodia}.
\newblock {\em Sustainability Accounting, Management and Policy Journal},
  10(3):451--475.

\bibitem[Wahl and Bull, 2014]{Wahl2014}
Wahl, A. and Bull, G.~Q. (2014).
\newblock Mapping research topics and theories in private regulation for
  sustainability in global value chains.
\newblock {\em Journal of Business Ethics}, 124(4):585--608.

\bibitem[Williams and Srnicek, 2013]{Williams2013}
Williams, A. and Srnicek, N. (2013).
\newblock \#accelerate manifesto for an accelerationist politics.
\newblock Critical Legal Thinking.

\bibitem[Wong, 2023]{Wong2023}
Wong, W.~H. (2023).
\newblock {\em We, the Data: Human Rights in the Digital Age}.
\newblock MIT Press.

\bibitem[{World Bank}, 2021]{WorldBank2021}
{World Bank} (2021).
\newblock World development report 2021: Data for better lives.
\newblock Technical report, World Bank.

\bibitem[Yang et~al., 2024]{Yang2024}
Yang, M., Guo, J., Zhu, L., Zhu, H., Song, X., Zhang, H., and Xu, T. (2024).
\newblock Fairness evaluation of marketing algorithms: a framework for equity
  distribution.
\newblock {\em Journal of Electronic Business \& Digital Economics},
  3(3):251--274.

\bibitem[Zuboff, 2015]{Zuboff2015}
Zuboff, S. (2015).
\newblock Big other: Surveillance capitalism and the prospects of an
  information civilization.
\newblock {\em Journal of Information Technology}, 30(1):75--89.

\bibitem[Zuboff, 2019]{Zuboff2019}
Zuboff, S. (2019).
\newblock {\em The Age of Surveillance Capitalism: The Fight for a Human Future
  at the New Frontier of Power}.
\newblock PublicAffairs.

\bibitem[Zuckerberg, 2017]{Zuckerberg2017}
Zuckerberg, M. (2017).
\newblock Building global community.
\newblock Facebook.

\end{thebibliography}

\newpage
\appendix
\section{Dignity-Centric Digital Social Contract. Operational Blueprint}\label{annex:blueprint}

While Section 3 defines the normative framework of a dignity-centric digital order, explaining the underlying rationale and clarifying the principles at stake and the reasons why dignity, rights, and democratic legitimacy must constrain data-driven technologies, this annex, by contrast, shifts from justification to implementation. It specifies how these core dimensions can be operationalised through institutional pillars, actionable governance principles, and enforceable non-negotiable limits, thereby transforming a declaratory vision into a concrete blueprint for practice across the data lifecycle.

\begin{figure}[htbp]
\centering
\includegraphics[width=0.9\textwidth]{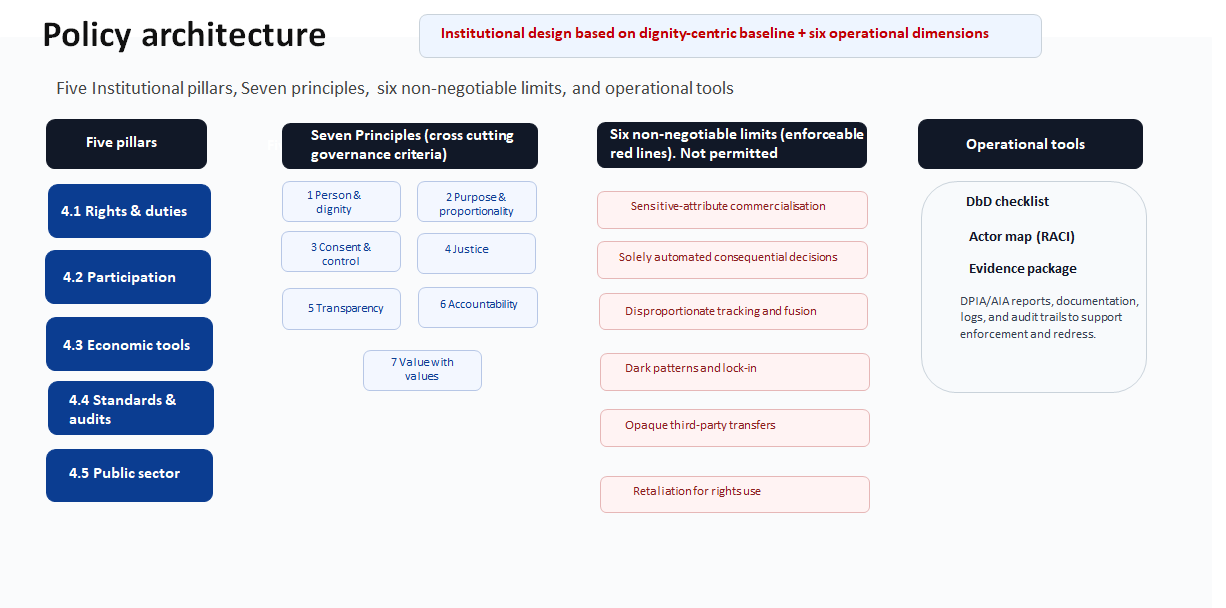}
\caption{Policy architecture of the dignity-centric digital contract.}
\label{fig:A1}
\end{figure}

In particular, we propose (Figure~\ref{fig:A1}) five institutional pillars (Rights and Duties, Participation and Collective Governance, Economic Tools, Technical Standards and Audits, Public Sector Leadership and Culture), seven principles (Primacy of the person and her dignity, Legitimate purpose and proportionality, Meaningful consent and continuous control, Non-discrimination and justice, Transparency and explainability, Responsibility and accountability, Value with values---shared benefit---) and six non-negotiable limits (Commercialisation of especially protected attributes, Fully automated consequential decisions without meaningful human intervention, Intrusive collection without necessity and minimisation (disproportionate tracking, surveillance, data fusion), Dark patterns and lock-in (manipulated consent and exit), Opaque transfers to third parties (hidden flows, no traceability), Retaliation for exercising rights (portability, erasure, objection, access). Finally, we offer some operational tools to reduce the gap between principle and practice, in order to help institutions adopt a minimal viable package that sequences implementation in a feasible order.

To clarify the relationship between the normative framework developed in Section 3 and the operational contract specified in this annex, Table~\ref{tab:A1} maps each of the six core dimensions onto the corresponding governance architecture. The table shows how broad conceptual commitments are translated into implementable institutional levers (five pillars), cross-cutting decision criteria (seven principles), and enforceable categorical constraints (six non-negotiable limits). This mapping highlights the coherence of the dignity-centric approach across levels of analysis, from foundational justification to practical governance mechanisms.

\subsection{Institutional Pillars}\label{sec:A1.1}

In this section we present five institutional pillars as implementable institutional levers to operationalize the core dimensions of the proposed dignity-centric digital social contract.

\paragraph{1. Rights and Duties.}
We begin with rights and duties because the proposed contract is rights-first. The paper treats personal data as dignity-relevant manifestations of the person while also recognising their productive role in contemporary value chains. Institutions should therefore protect a non-waivable baseline of digital rights and should assign affirmative duties to controllers and processors.

\begin{itemize}
\item Enshrine dignity-centric digital rights, recognizing personal data as extensions of personality and, in some cases, of community identity.
\item Impose duties of loyalty, care, and confidentiality on data controllers and processors.
\item Prohibit inherently dignity-eroding practices (e.g., unconstrained sale of sensitive data; fully automated consequential decisions without human oversight).
\end{itemize}

\paragraph{2. Participation and Collective Governance.}
\begin{itemize}
\item Enable data cooperatives and trade unions as vehicles for collective bargaining and shared benefit.
\item Support trusted data intermediaries, such as those regulated under the Data Governance Act and similar frameworks.
\item Require participatory governance for civic data spaces (health, urban, environmental).
\end{itemize}

\paragraph{3. Economic Tools.}
\begin{itemize}
\item Promote Data Agents that enable citizens to recover their data quickly and easily. By Data Agents we mean user-facing technical and organisational services (public, private, or cooperative) that execute portability, access, rectification, and deletion requests on behalf of individuals, with verifiable logs and non-retaliation safeguards.
\item Create new business models based on transparent data marketplaces.
\item Pilot data dividends and fair licensing for selected use cases, ensuring they complement---rather than replace---rights-based safeguards.
\item Implement Pigouvian informational taxes on high-risk profiling and sensitive inference, with revenues channelled to affected groups and public goods.
\item Update competition law to treat data accumulation and behavioural lock-in as central concerns.
\end{itemize}

\paragraph{4. Technical Standards and Audits.}
\begin{itemize}
\item Operationalize DbD via standards bodies (W3C, ISO, IEEE).
\item Expand the mandate of accredited trustees so that they certify data integrity and act as custodians of such data. In this sense, central banks represent one possible model where mandates already support integrity and audit capacity.
\item Require AIAs, DPIAs, and periodic audits for high-impact systems.
\item Standardize documentation (data sheets, model cards, system cards).
\end{itemize}

\paragraph{5. Public-Sector Leadership and Culture.}
\begin{itemize}
\item Use public procurement to demand DbD compliance.
\item Promote data and AI literacy in education systems and public-sector training.
\item Support research and ``living labs'' that align data use with the common good.
\end{itemize}

The five pillars identify institutional levers, but they do not provide decision criteria for specific systems, products, and data practices. Sections~\ref{sec:A1.2} and~\ref{sec:A1.3} supply these criteria.

\subsection{Seven Principles}\label{sec:A1.2}

In this section we propose seven guiding principles for DbD deployments. These seven principles function as general governance tests that should apply across sectors and across the data lifecycle.

\begin{enumerate}
\item \textbf{Primacy of the person and her dignity.} Individuals are the primary right-holders in relation to the processing of their personal data. Personal data are dignity-relevant manifestations of the person and must not be treated as mere merchandise or fuel. Data sovereignty should be understood here not as full ownership of an asset, but as enforceable positions of control and protection over the data lifecycle: meaningful access, rectification, erasure where appropriate, objection and restriction, portability, and strong limits on high-impact profiling and automated consequential decisions.

\item \textbf{Legitimate purpose and proportionality.} Only data processing that is necessary and proportionate to just, explicit purposes is allowed.

\item \textbf{Meaningful consent and continuous control.} Consent must be easy to give, granular, revocable without friction, and free from dark patterns or coercion.

\item \textbf{Non-discrimination and justice.} Systems must undergo impact assessments and bias testing, including intersectional harms, with concrete mitigation measures.

\item \textbf{Transparency and explainability.} Systems must be understandable to users and auditable by third parties; explanations should be tailored to non-experts.

\item \textbf{Responsibility and accountability.} There must be clear allocation of responsibilities, robust audit trails, accessible remedies, and effective sanctions for violations.

\item \textbf{``Value with values'' (shared benefit).} Fair redistribution of the value generated by data, including support for public interest uses under strong safeguards.
\end{enumerate}

\subsection{Six Non-Negotiable Limits}\label{sec:A1.3}

The six non-negotiable limits state categorical prohibitions that remain in force even when organisations claim user consent, since as discussed, consent often operates under dependency, opacity, and asymmetric bargaining power (Section 2.5). They operationalise the dignity-centric baseline defended in Section 2 and translate it into enforceable constraints. Each red line is anchored in (i) one or more guiding principles and (ii) concrete legal-institutional instruments (ex ante impact assessments, prohibitions, governance duties, auditability and remedies).

\begin{enumerate}
\item \textbf{Commercialisation of especially protected attributes.}

\textit{Principle anchor:} Primacy of the person and her dignity; Non-discrimination and justice; Legitimate purpose and proportionality.

\textit{Instrument anchor:} (a) sectoral and contextual prohibitions on targeting/profiling based on sensitive attributes; (b) mandatory DPIA for sensitive-inference pipelines and ad-tech profiling; (c) auditing and bias-testing duties for systems that can infer protected traits from proxies; (d) fiduciary-like duties of loyalty for platforms and AI providers toward users' data interests (to prevent dignity-eroding exploitation even where formal consent is obtained).

\item \textbf{Fully automated consequential decisions without meaningful human intervention.}

\textit{Principle anchor:} Responsibility and accountability; Transparency and explainability; Primacy of dignity.

\textit{Instrument anchor:} (a) strict application of GDPR Art.\ 22 safeguards: no solely automated decisions with legal or similarly significant effects without meaningful human oversight, contestability and explanation; (b) AIAs/AIA templates for high-impact decision systems; (c) periodic audits (internal/external) and traceability (model cards/system cards) to evidence compliance.

\item \textbf{Intrusive collection without necessity and minimisation (disproportionate tracking, surveillance, data fusion).}

\textit{Principle anchor:} Legitimate purpose and proportionality; Responsibility and accountability; Meaningful consent and continuous control.

\textit{Instrument anchor:} (a) enforce purpose limitation and minimisation as operational requirements (data inventories, retention limits, prohibition of unnecessary identifiers); (b) DPIA as default for large-scale tracking and data-fusion architectures; (c) technical PETs (e.g., differential privacy, federated learning) and security controls as ex ante constraints; (d) competition and platform rules treating behavioural lock-in and data accumulation as regulated risk factors.

\item \textbf{Dark patterns and lock-in (manipulated consent and exit).}

\textit{Principle anchor:} Meaningful consent and continuous control; Transparency and explainability; Respect for autonomy (as stated earlier in the ethical commitments).

\textit{Instrument anchor:} (a) explicit prohibition of manipulative interface patterns and coercive consent flows, backed by usability standards and regulator-ready evidence; (b) A/B testing governance and ``choice architecture'' review as part of DPIA/AIA; (c) DSA-aligned obligations on transparency of recommender/ads interfaces and bans on dark patterns; (d) remedies and sanctions tied to systematic design-based coercion.

\item \textbf{Opaque transfers to third parties (hidden flows, no traceability).}

\textit{Principle anchor:} Transparency and explainability; Responsibility and accountability; Legitimate purpose and proportionality.

\textit{Instrument anchor:} (a) mandatory end-to-end traceability across the chain (records of processing, data-sharing registers, versioning); (b) clear contractual allocation of roles and liabilities (controller/processor/joint controllership) and audit rights across partners; (c) vetted transfer mechanisms and documented third-party due diligence; (d) trusted intermediaries and neutral data intermediation where relevant (DGA logic) to reduce opaque sharing.

\item \textbf{Retaliation for exercising rights (portability, erasure, objection, access).}

\textit{Principle anchor:} Responsibility and accountability; Primacy of the person and her dignity; ``Value with values'' (shared benefit) insofar as basic participation must not be conditioned on surrendering rights.

\textit{Instrument anchor:} (a) explicit non-retaliation rules in product governance and contractual terms; (b) complaint-handling and remedy pathways with due-process guarantees (internal escalation, independent ombudsperson, collective redress); (c) monitoring and enforcement metrics (drop in service quality, price discrimination, access degradation) as compliance indicators; (d) portability tooling (Data Agents) as a practical guarantee that rights are exercisable without penalty.
\end{enumerate}

\subsection{Operational Tools}\label{sec:A1.4}

Beyond institutional pillars, governance principles, and categorical red lines, a dignity-centric contract must also confront a persistent gap between normative commitment and operational practice. Principles alone risk remaining declaratory unless they are translated into concrete organisational routines, verifiable artefacts, and enforceable supervision mechanisms. For this reason, the contract is complemented by a set of practical governance tools that enable dignity-by-design to function not only as an ethical orientation but as an auditable and implementable compliance architecture across the full data lifecycle.

Thus, to ensure that the contract does not remain at the level of abstract principles, we introduce an operational toolkit designed to support implementation, accountability, and enforcement. First, the Dignity-by-Design checklist that we present provides a minimum control set for system design and deployment, applicable across procurement, engineering workflows, compliance processes, and external audits.

For each product, service, or AI system, teams should verify at least ten controls: purpose and proportionality; data inventory and minimization; legal basis/consent; portability processes; bias and impact assessment; user-facing explanations; versioning and traceability; PETs; internal/external audits; and remedy mechanisms with guarantees of non-retaliation.

By requiring teams to verify these core controls the checklist translates normative requirements into concrete and testable obligations. Importantly, these controls scale with risk: as processing moves from raw traces to profiling and consequential inference, duties of minimisation, explanation, auditability, and redress become correspondingly stricter.

Second, we propose creating an actor map, a matrix of actors $\times$ obligations that clarifies responsibility across the data ecosystem by assigning obligations along the full chain of actors, from individuals and service providers to aggregators, ad networks, partners, and regulators. This governance matrix operationalises accountability by specifying who is responsible, accountable, consulted, and informed for each key duty, thereby reducing ambiguity and preventing responsibility from being diffused across complex data infrastructures.

Finally, the contract requires enforcement and remedy pathways that work both for individual complaints and for systemic harms. Organisations should implement structured escalation procedures, including independent internal review, access to external dispute-resolution bodies or ombudspersons, and collective redress mechanisms where population-level profiling produces shared harms. For contested automated decisions, sufficient documentation must be preserved to enable meaningful review, including the data sources, legal basis, model version, human oversight role, explanations provided, and linked impact assessments. Regulators, in turn, should tie sanctions to repeated design-based violations, with escalation from corrective orders and intensified audits to fines, certification withdrawal, procurement exclusion, and, where necessary, temporary suspension of unlawful processing until redesign and verification occur.

Together, these operational instruments close the gap between principle and practice, ensuring that dignity-by-design is not merely aspirational but institutionally actionable, auditable, and enforceable.

\begin{landscape}
\begin{table}[p]
\centering
\small
\caption{Mapping the Six Core Dimensions of the Dignity-Centric Framework to Their Operationalisation Through Institutional Pillars, Governance Principles, and Non-Negotiable Limits.}
\label{tab:A1}
\begin{tabular}{p{3.5cm}p{5.5cm}p{5.5cm}p{6.5cm}}
\toprule
\textbf{Core Dimension} & \textbf{Institutional Pillars (Implementable Levers)} & \textbf{Governance Principles (Decision Criteria)} & \textbf{Non-Negotiable Limits (Categorical Constraints)} \\
\midrule
1.\ Technological Design and Governance &
Technical Standards and Audits; Rights and Duties; Public Sector Leadership &
Transparency and explainability; Responsibility and accountability; Legitimate purpose and proportionality &
Fully automated consequential decisions without meaningful human intervention; Intrusive collection without necessity and minimisation; Opaque transfers to third parties \\
\addlinespace
2.\ Human Oversight and Limits to Automation &
Rights and Duties; Technical Standards and Audits &
Primacy of the person and her dignity; Responsibility and accountability; Transparency and explainability &
Fully automated consequential decisions without meaningful human intervention; Retaliation for exercising rights \\
\addlinespace
3.\ Value, Redistribution, and Incentives &
Economic Tools; Participation and Collective Governance; Rights and Duties &
``Value with values'' (shared benefit); Non-discrimination and justice; Primacy of dignity &
Commercialisation of especially protected attributes; Retaliation for exercising rights \\
\addlinespace
4.\ Legitimacy and Multi-Actor Governance &
Participation and Collective Governance; Rights and Duties; Public Sector Leadership &
Responsibility and accountability; Transparency and explainability; Meaningful consent and continuous control &
Opaque transfers to third parties; Dark patterns and lock-in \\
\addlinespace
5.\ Sociocultural Cohesion and the Digital Commons &
Public Sector Leadership and Culture; Participation and Collective Governance &
Primacy of dignity; Non-discrimination and justice; Meaningful consent and continuous control &
Commercialisation of especially protected attributes; Dark patterns and lock-in \\
\addlinespace
6.\ Legal-Regulatory Guarantees and Enforceability &
Rights and Duties; Technical Standards and Audits; Public Sector Leadership &
Responsibility and accountability; Legitimate purpose and proportionality; Transparency and explainability &
All six limits function as enforceable floor; particularly intrusive collection, opaque transfers, automated decisions \\
\bottomrule
\end{tabular}
\end{table}
\end{landscape}

\section{Glossary}\label{annex:glossary}

\paragraph{Personal data.}
Any information relating to an identified or identifiable natural person, where identifiability includes direct or indirect identification by reference to identifiers (name, ID number, location data, online identifier, etc.) or factors specific to identity. (Regulation (EU) 2016/679, GDPR, Art.\ 4(1), and identifiability test in Recital 26; Council of Europe, Convention ETS No.\ 108, Art.\ 2(a); Charter of Fundamental Rights of the EU, Art.\ 8).

\paragraph{Inferred / derived data.}
Personal data generated by processing other data (provided or observed) through analysis, including statistical modelling or machine learning, producing new attributes, scores, predictions, or profiles not directly supplied by the data subject. (GDPR: profiling as analysis/prediction of personal aspects, Art.\ 4(4); Recital 30 on online identifiers leaving traces used to create profiles; Recital 71 on profiling and assumptions from analysing sets of data. WP29/EDPB typology: data provided, observed, and inferred/derived is set out in Guidelines WP251rev.01 Article 29 Working Party, Guidelines on Automated Individual Decision-making and Profiling (WP251rev.01), adopted 6 February 2018, endorsed by the EDPB.

Regulatory consequence for high-impact inferential uses: AI systems used in certain high-stakes domains listed in Annex III (of the European Union AI Regulation) are treated as high-risk when they fall within the Regulation's high-risk classification rules, triggering the Title III compliance regime (including information/transparency obligations, human oversight, and conformity assessment requirements). (Regulation (EU) 2024/1689, AI Act, notably Art.\ 6 (linking to Annex III) and Title III; Annex III; plus transparency/human oversight and conformity assessment duties within Title III.)

\paragraph{Relational / co-produced data.}
Personal data that are generated by, or reveal, patterns of interaction, association, or co-presence among two or more persons (e.g., communication and social-graph data, household or co-location traces), such that the information cannot be attributed exclusively to one data subject and may affect multiple persons simultaneously. (Anchored in the breadth of personal data as information relating to any identifiable natural person: GDPR Art.\ 4(1), Recital 26; in communications confidentiality protecting all parties: ePrivacy Directive 2002/58/EC, Art.\ 5(1); and in EU law's recognition of shared responsibility in data ecosystems via joint controllership: GDPR Art.\ 26, and CJEU case law such as Case C-210/16, \textit{Wirtschaftsakademie} (joint controllership in platform interaction contexts).)

\paragraph{Data sovereignty.}
A multidimensional governance construct designating the rights, capacities, and institutional arrangements through which individuals (and, where relevant, communities and public authorities) exercise meaningful and effective authority over the collection, processing, access to, use of, sharing of, and value creation from data---through: (a) protection (fundamental-rights safeguards), (b) participation (agency and accountable governance), and (c) provision (technical/organisational/legal-operational conditions that make rights effective in practice). (UN GA Resolution A/RES/68/167 (2013) frames privacy protections in the digital age as a human-rights compliance demand; UN GA adopted the Global Digital Compact as Annex I to A/RES/79/1 (22 September 2024), an intergovernmental framework for digital cooperation (including data governance commitments); at EU level, the triptych GDPR (rights protection), Data Governance Act (Regulation (EU) 2022/868) (data governance mechanisms such as intermediaries/altruism/trust), and the Data Act (Regulation (EU) 2023/2854) (access/portability and data-use rules) operationalises key pillars.)

\paragraph{Human dignity (conception used).}
Human dignity is treated as the inviolable, inherent worth of every person, functioning in this paper as: (i) a foundational value grounding privacy and data protection as rights (not merely interests), and (ii) a particularly weighty constitutional interest that can justify strong constraints on certain practices---while remaining within the general structure of legal justification and review in EU law. (Charter of Fundamental Rights of the EU, Art.\ 1: ``Human dignity is inviolable. It must be respected and protected''; CJEU, Case C-36/02, \textit{Omega}, recognising dignity as a legitimate public-policy ground capable of justifying restrictions; EDPS, Opinion 4/2015, calling for dignity at the heart of a new digital ethics and linking dignity to privacy/data protection; UNESCO Recommendation on the Ethics of AI (2021), Section III (Values), placing respect for human rights and fundamental freedoms and human dignity at the foundation of AI governance.)

Within this framework, a further conceptual clarification is useful. \textit{Individual dignity} refers to the subject as a bearer of rights and obligations with a unique identity---that is, the individual considered as an element within a legal and institutional system. \textit{Personal dignity}, by contrast, incorporates the relational dimension of the moral subject: not only the network of relationships that generate rights and duties, but also those grounded in shared and transcendent values that extend beyond purely functional or institutional roles.

Thus, human dignity operates at the foundational level, expressing the inherent and equal worth of every human being. Individual dignity represents the juridical and institutional expression of that worth, focusing on the individual as a rights-holder within a structured legal order. Personal dignity deepens the account by emphasising the relational and moral constitution of the person, highlighting dimensions of identity, responsibility, and community that cannot be reduced to formal rights alone. Together, they provide a layered understanding of dignity that grounds legal protection while acknowledging the broader moral and social conditions under which human beings flourish.

\paragraph{Human disruption.} Following \citet{Arendt1973}, we can say that human beings in society are characterized by their capacity to act as unique individuals and to participate in a shared world. In this paper, we use the term \textit{human disruption} to refer to the impacts that technological disruption may provoke on the status quo of the human being in society, including: (a) impacts on the size and depth of the public sphere, and (b) impacts on the human capacity to appear and act within it as a moral and relational subject.

\subsection*{Sources}

\begin{itemize}[leftmargin=*]
\item Arendt, H. (1973). \textit{The origins of totalitarianism} (Vol.\ 244). Houghton Mifflin Harcourt.

\item Charter of fundamental rights of the European Union, article 8.\\
\url{https://eur-lex.europa.eu/legal-content/EN/TXT/?uri=oj:JOC_2000_364_R_0001_01}

\item Council of Europe, Convention for the Protection of Individuals with regard to Automatic Processing of Personal Data (ETS No.\ 108), Art.\ 2(a).\\
\url{https://rm.coe.int/1680078b37}

\item Court of Justice of the European Union, Case C-36/02, \textit{Omega Spielhallen- und Automatenaufstellungs-GmbH v Oberb\"{u}rgermeisterin der Bundesstadt Bonn}, Judgment of 14 October 2004, ECLI:EU:C:2004:614.\\
\url{https://eur-lex.europa.eu/legal-content/EN/TXT/?uri=CELEX:62002CJ0036}

\item Court of Justice of the European Union, Case C-210/16, \textit{Unabh\"{a}ngiges Landeszentrum f\"{u}r Datenschutz Schleswig-Holstein v Wirtschaftsakademie Schleswig-Holstein GmbH}, Judgment of 5 June 2018, ECLI:EU:C:2018:388.\\
\url{https://eur-lex.europa.eu/legal-content/EN/TXT/?uri=CELEX:62016CJ0210}

\item Directive 2002/58/EC (ePrivacy Directive), Art.\ 5(1).\\
\url{https://eur-lex.europa.eu/eli/dir/2002/58/oj/eng}

\item European Commission (Article 29 Working Party newsroom redirection), Guidelines on Automated individual decision-making and Profiling for the purposes of Regulation (EU) 2016/679 (WP251rev.01).\\
\url{https://ec.europa.eu/newsroom/article29/redirection/item/612053}

\item European Data Protection Board, Endorsed WP29 Guidelines list (including WP251rev.01).\\
\url{https://www.edpb.europa.eu/our-work-tools/general-guidance/endorsed-wp29-guidelines_en}

\item European Data Protection Supervisor, Opinion 4/2015, Towards a New Digital Ethics: Data, dignity and technology (11 September 2015).\\
\url{https://www.edps.europa.eu/sites/default/files/publication/15-09-11_data_ethics_en.pdf}

\item Regulation (EU) 2016/679 (General Data Protection Regulation), Arts.\ 4(1), 4(4), 5(1)(d), 13--14, 21--22, 26; Recitals 26, 30, 71.\\
\url{https://eur-lex.europa.eu/eli/reg/2016/679/oj/eng}

\item Regulation (EU) 2022/868 (Data Governance Act).\\
\url{https://eur-lex.europa.eu/eli/reg/2022/868/oj/eng}

\item Regulation (EU) 2023/2854 (Data Act).\\
\url{https://eur-lex.europa.eu/eli/reg/2023/2854/oj/eng}

\item Regulation (EU) 2024/1689 (AI Act) (incl.\ Art.\ 6, Annex III; Title III).\\
\url{https://eur-lex.europa.eu/eli/reg/2024/1689/oj/eng}

\item UNESCO, Recommendation on the Ethics of Artificial Intelligence (2021) (adopted 23 November 2021).\\
\url{https://unesdoc.unesco.org/ark:/48223/pf0000380455}

\item United Nations General Assembly, Resolution A/RES/68/167 (2013), The right to privacy in the digital age.\\
\url{https://digitallibrary.un.org/record/764407/files/A_RES_68_167-EN.pdf}

\item United Nations General Assembly, Resolution A/RES/79/1 (22 September 2024), Annex I: Global Digital Compact.\\
\url{https://docs.un.org/en/a/res/79/1}
\end{itemize}

\end{document}